\documentclass[reprint,aps,pra,superscriptaddress,longbibliography,nofootinbib,floatfix,10pt]{revtex4-2}

\usepackage[T1]{fontenc}
\usepackage{amsmath,amssymb,mathtools,bm,bbm}
\usepackage[dvipsnames]{xcolor}
\usepackage[colorlinks=true,linkcolor=MidnightBlue,citecolor=MidnightBlue,urlcolor=MidnightBlue]{hyperref}
\usepackage{graphicx}
\usepackage{booktabs}
\usepackage{array}
\usepackage{multirow}
\usepackage{enumitem}
\usepackage{qcircuit}
\usepackage{placeins}

\newcommand{\id}{\hat{\mathbb I}}
\newcommand{\Hhat}{\hat H}
\newcommand{\rhoh}{\hat\varrho}
\newcommand{\avg}[1]{\left\langle #1 \right\rangle}
\newcommand{\ket}[1]{\left|#1\right\rangle}
\newcommand{\bra}[1]{\left\langle #1\right|}
\newcommand{\abs}[1]{\left|#1\right|}

\newcommand{\rv}{\bm r}
\newcommand{\sigj}[1]{\hat\sigma^{(j)}_{#1}}
\newcommand{\sigs}[1]{\hat\sigma^{(s)}_{#1}}
\newcommand{\svec}[1]{\hat{\bm\sigma}^{(#1)}}
\newcommand{\calC}{\mathcal{C}}
\newcommand{\calN}{\mathcal{N}}

\newcommand{\calG}{\mathcal{G}}

\newcommand{\Fglob}{F_{\mathrm{glob}}}
\newcommand{\Fdec}{F_{\mathrm{dec}}}

\DeclareMathOperator{\Var}{Var}
\DeclareMathOperator{\Tr}{Tr}

\begin{document}

\title{Operator spreading and recoverability of local quantum Fisher information in a $U(1)$-broken spin chain}%\\[1.5em]

\author{Marcin P{\l}odzie\'n}
\affiliation{Qilimanjaro Quantum Tech, Carrer de Vene\c{c}uela 74, 08019 Barcelona, Spain}

\author{Jan Chwede\'nczuk}
\affiliation{Faculty of Physics, University of Warsaw, ulica Pasteura 5, 02-093 Warszawa, Poland}

\begin{abstract}
  While out-of-time-order correlators establish a causal light cone for operator spreading, they do not guarantee that the parameter sensitivity carried by the operator remains locally recoverable. We examine the distinction between operator spreading and metrological recoverability for a parameter encoded in a single site of an XX spin chain subjected to a $U(1)$-breaking transverse field. 
  We evaluate three levels of local metrological accessibility: the bare single-site quantum Fisher information (QFI), 
  the QFI recovered by a variational sweep decoder acting on a finite spatial block, and the exact block QFI. 
  In the integrable limit, the sensitivity propagates as a one-magnon wave packet, and a single-qubit decoder recovers the full block QFI.
  Breaking magnon-number conservation couples the parameter tangent state to multi-magnon sectors. We analytically demonstrate that the local QFI has no first-order correction in field strength; the leading depletion enters at $\mathcal{O}(h^2)$ through two-magnon scattering.
  As the field strength increases, the decoded QFI falls below the exact block QFI---a gap reflecting a generic finite-dimensional compression limitation, as a single output qubit generically cannot capture the full QFI of a block state whose parameter dependence spans more than an effective two-dimensional subspace. 
  The block QFI itself falls below the conserved global value, confirming that the sensitivity has spread beyond the block into non-local correlations. 
  This operational hierarchy provides a precise quantitative distinction between the arrival of operator support and the local accessibility of metrological information.
\end{abstract}

\maketitle

\section{Introduction}
\label{sec:intro}
In quantum many-body systems, the propagation of local perturbations is typically characterized by out-of-time-order correlators (OTOCs), which diagnose the growth of operator support and the emergence of a dynamical light cone~\cite{Swingle2018,XuSwingle2024,Swingle2016,Garttner2017,XuSwingle2020,LiebRobinson1972}. 
Quantum Fisher information (QFI) quantifies a state's sensitivity to an encoded parameter and is 
widely used for metrological and entanglement diagnostics~\cite{BraunsteinCaves1994,Hyllus2012,Toth2012,Pezze2018}. While OTOCs track this operator spreading, QFI quantifies the distinguishability of nearby parameter-encoded states under an allowed measurement. However, operator support can arrive without bringing recoverable metrological information.
We study how metrological information is redistributed for a parameter encoded on a single site of a spin chain that evolves under a parameter-independent Hamiltonian. Since unitary evolution conserves the total QFI, the relevant question is not whether information is lost globally, but rather, how it is distributed across space and correlations. 

We compare three measures of local accessibility at a distant probe site. The first is the bare single-site QFI, which is directly available by local measurement. The second is the decoded QFI, which is concentrated onto that site by optimizing a unitary on a finite spatial block around the probe. The third is the exact block QFI, which is the total sensitivity present in that neighborhood before any decoding. Their separation quantifies the scale at which sensitivity becomes inaccessible. When the dynamics are integrable, the encoded sensitivity propagates as a single-particle excitation. In this case, the block QFI equals the total single-particle weight within the block. A local unitary can then concentrate this weight entirely onto a single output site. Although sensitivity spreads spatially, it remains locally recoverable. The bare probe QFI may be small simply because the excitation has not yet arrived; it is not because information has been scrambled.

Perturbations that break magnon-number conservation couple the one-magnon tangent state to the vacuum and higher-magnon sectors.
The sensitivity concentrated in a single excitation is then redistributed into multi-magnon amplitudes and correlations.
At weak coupling, this redistribution can be localized enough that a sufficiently wide block decoder can recover much of the sensitivity.
At stronger coupling, however, the chosen block and decoder no longer capture the full reduced-state distinguishability.
The hierarchy identifies whether the missing sensitivity is present in the block but not extracted by the decoder or absent from the block altogether.

We demonstrate that the summed squared commutator provides an upper bound on local response. 
Once the perturbed dynamics generate multi-body correlations, this bound becomes a poor predictor of the extractable sensitivity. 
The squared commutator can reach its finite-system ceiling, indicating that operator support has arrived at the probe~\cite{YoshidaYao2019}, 
while the locally recoverable QFI can decrease because the parameter sensitivity is distributed over many-body correlations~\cite{LewisSwan2019,Garttner2018}.

The central result of this work is the operational distinction between operator spreading and metrological recoverability. While the squared commutator provides a causal upper bound on the local response, it does not determine whether the sensitivity carried by the spreading operator can be extracted through a finite block operation. 
The hierarchy of bare probe QFI, sweep-decoded QFI, exact block QFI, and conserved global QFI resolves this distinction. In the XX limit, the hierarchy collapses within each block because the tangent state remains a one-magnon wave packet, and a shallow sweep decoder concentrates all block sensitivity onto the output qubit. However, as the transverse field increases, these quantities diverge. The bare probe QFI can be much smaller than the locally decoded signal. The decoded QFI can fall below the exact block QFI. And the block QFI itself can fall below the conserved global value.

The remainder of the paper is organized as follows. Section~\ref{sec:setup} introduces the metrological setup and the OTOC-based bound. Section~\ref{sec:analytical_xx} presents the exact XX benchmark. Section~\ref{sec:perturbation} analyzes the $U(1)$-breaking perturbation. Section~\ref{sec:decoder} studies the variational block decoder across the resulting crossover. Section~\ref{sec:discussion} summarizes the results and their implications.

% ═══════════════════════════════════════════════════════════════
%  II. SETUP
% ═══════════════════════════════════════════════════════════════

\section{Setup}\label{sec:setup}

We consider $N$ spin-$1/2$ degrees of freedom on a one-dimensional chain with open boundary conditions. 
We initialize the system in the fully polarized reference state being a product of $N$ eigenstates of $\hat\sigma_z^{(i)}$ operators, $\ket{\psi_0}=\ket{\uparrow}_z^{\otimes N}$. 
This is a zero-magnon state that is an eigenstate of some magnon-preserving Hamiltonian $\hat H_0$. 
A metrological parameter $\theta$ is encoded on a single source site $s$ through the localized unitary rotation $\hat U_s(\theta)=e^{-i\theta\sigs{y}/2}$, giving the initial state
\begin{equation}
  \ket{\psi_{\theta,0}} = \hat U_s(\theta) \ket{\psi_0},
\end{equation}
whose tangent vector at $\theta=0$ is a single localized magnon. 
We focus on the null-signal limit ($\theta=0$) representing weak-signal quantum sensing, where the encoding acts as an infinitesimal perturbation to the vacuum.
The state subsequently evolves under a parameter-independent many-body Hamiltonian $\Hhat(h)$
\begin{equation}
  \ket{\psi_\theta(t)} = e^{-i\Hhat(h) t} \ket{\psi_{\theta,0}}.
\end{equation}
The target $U(1)$-breaking Hamiltonian consists of two parts, namely
\begin{align}
  \Hhat(h) = \Hhat_0 +h \hat V,
\end{align}
where $\hat V = \sum_i \hat\sigma_i^x$ is the symmetry-breaking perturbation.
For $h>0$ the polarized state is no longer an eigenstate,
and its spectral decomposition involves multiple magnon-number sectors.

\begin{figure}[t!]
\centering
\includegraphics[width=\columnwidth]{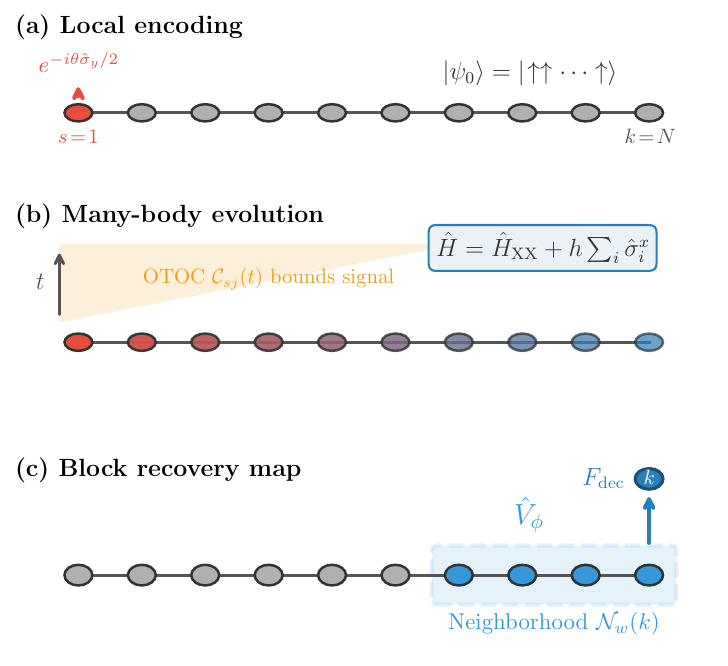}
\caption{\textbf{(a)} A local unitary encoding injects one unit of quantum Fisher information ($F_\mathrm{glob}=1$) at the source. \textbf{(b)} The parameter sensitivity spreads under $\hat{H}(h)$, bounded by the OTOC lightcone. At $h=0$, information transports as a locally recoverable single-particle excitation. For $h>0$, the initial excitation scatters into many-body correlations. \textbf{(c)} A variational unitary $\hat{V}_\phi$ on a block of width $w$ concentrates the block QFI onto an output qubit ($F_{\mathrm{dec}}$). The inequality $F_{\mathrm{dec}} < F_Q(\hat\varrho_{\mathcal{N}_w})$ identifies sensitivity present in the block but not concentrated by the chosen sweep decoder, while $F_Q(\hat\varrho_{\mathcal{N}_w}) < F_\mathrm{glob}$ indicates that the chosen block alone no longer contains the full global distinguishability.}
\label{fig:schematic}
\end{figure}

Unitary evolution of a pure state conserves the global quantum Fisher information evaluated across the entire chain. It is determined by the variance of the rotation generator evaluated on the initial state
\begin{equation}
\Fglob \equiv F_Q\!\big(\ket{\psi_\theta(t)}\big) = 4\Var_{\psi_0}\!\left(\frac{\sigs{y}}{2}\right).
\end{equation}
For our polarized initial state $\ket{\psi_0}$, the local expectation value $\avg{\sigs{y}}_0$ vanishes, yielding $\Fglob=1$. Therefore, at $\theta=0$, the metrological tangent vector is a single localized spin flip carrying one unit of global QFI at the source site $s$.

Single-qubit measurements at probe site $j$ yield the reduced state
\begin{equation}\label{eq.red}
\rhoh_j(\theta,t)=\Tr_{\overline{j}}\!\big[\rhoh_\theta(t)\big]
=\frac12\big(\id+\rv_j(\theta,t)\cdot\svec{j}\big),
\end{equation}
where $r_j^\alpha(\theta,t)=\avg{\sigj{\alpha}(t)}_{\theta,0}$ and $\abs{\rv_j}\le 1$. The qubit QFI with respect to $\theta$ can be written in Bloch-vector form~\cite{BraunsteinCaves1994,Paris2009,Liu2020JPA} as
\begin{equation}
\label{eq:blochqfi}
F_j=\abs{\partial_\theta\rv_j}^2
+\frac{(\rv_j\cdot\partial_\theta\rv_j)^2}{1-\abs{\rv_j}^2},
\end{equation}
with the pure-state limit $\abs{\rv_j}\to 1$ understood by continuity when $\rv_j\cdot\partial_\theta\rv_j=0$~\cite{BraunsteinCaves1994}. Unless otherwise stated, the QFI is evaluated at the operating point $\theta=0$.

To bound the metrological response, we rely on a local linear-response identity.
When the parameter is encoded by the unitary $\hat U_s(\theta)=e^{-i\theta \hat\sigma_s^y/2}$,
the $\theta$-derivative of the initial encoded state is
$\partial_\theta \hat\varrho_{\theta,0} = -\frac{i}{2}[\hat\sigma_s^y, \hat\varrho_{\theta,0}]$.
Consequently, the parameter sensitivity of any local Pauli observable $\hat\sigma_j^\alpha$ evolves as
\begin{align}
  \partial_\theta r_j^\alpha(\theta,t) &= \mathrm{Tr}\!\left[\hat\sigma_j^\alpha(t) \partial_\theta \varrho_{\theta,0}\right]= \frac{i}{2} \langle [\hat\sigma_s^y, \hat\sigma_j^\alpha(t)] \rangle_{\theta,0}.
\end{align}
Applying the Cauchy--Schwarz inequality yields an upper bound in terms of the component squared commutators
\begin{equation}
4|\partial_\theta r_j^\alpha|^2 \le \left\langle [\hat\sigma_s^y,\hat\sigma_j^\alpha(t)]^\dagger [\hat\sigma_s^y,\hat\sigma_j^\alpha(t)] \right\rangle_{\theta,0} \equiv C_{sj}^{(\alpha)}(t;\theta).
\end{equation}
Summing over the three Pauli components $\alpha \in \{x,y,z\}$, we introduce the summed local OTOC norm
\begin{equation}
\mathcal{C}_{sj}^{\mathrm{sum}}(t;\theta) = \sum_{\alpha=x,y,z} C_{sj}^{(\alpha)}(t;\theta).
\end{equation}
This establishes the bound $4|\partial_\theta \mathbf{r}_j|^2 \le \mathcal{C}_{sj}^{\mathrm{sum}}(t;\theta)$. 

Combining this bound with the Bloch-sphere QFI formula (derived in Appendix~\ref{app:bloch}) establishes a hierarchy for the single-qubit reduced state from Eq.~\eqref{eq.red}
\begin{equation}
\label{eq:qfi_hierarchy}
(1-|\mathbf{r}_j|^2)F_j(t;\theta) \le |\partial_\theta \mathbf{r}_j|^2 \le F_j(t;\theta) \le \frac{\mathcal{C}_{sj}^{\mathrm{sum}}(t;\theta)}{4(1-|\mathbf{r}_j|^2)}.
\end{equation}
This hierarchy connects metrological sensitivity to entanglement generation. The denominator penalty $1-|\mathbf{r}_j|^2$ defines the mixedness of the local probe; for globally pure evolution, this quantity equals twice the linear entropy $S_L = 1 - \mathrm{Tr}\left[\rhoh_j^2\right]$, reflecting the bipartite entanglement between the output qubit and the rest of the chain. For nearly pure local reduced states the final upper bound can become weak or singular inside the light cone. Its robust content is causal: outside the Lieb--Robinson light cone the commutator norm is exponentially small, and therefore so is the local response. Before the operator front arrives, the numerator $\mathcal{C}_{sj}^{\mathrm{sum}}$ and the denominator penalty $(1-|\mathbf{r}_j|^2)$ both vanish identically, maintaining a trivial $0 \le 0$ bound; the singularity inside the light cone for pure states should be understood by continuity. For pure local states ($|\mathbf{r}_j| = 1$), $\mathbf{r}_j \cdot \partial_\theta \mathbf{r}_j = 0$ and the QFI reduces to the tangent norm $F_j = |\partial_\theta \mathbf{r}_j|^2$. As number-nonconserving dynamics generate multi-body correlations,
the probe can trace out to a mixed state. In that regime the OTOC upper bound becomes loose: the commutator can saturate at its finite-system value while the local QFI continues to decay, making the bound uninformative for the magnitude of the sensitivity. This looseness inside the lightcone is precisely why the operational decoder hierarchy introduced in Sec.~\ref{sec:decoder} is needed.

Because the commutator norm is constrained by Lieb--Robinson causality~\cite{LiebRobinson1972,RobertsSwingle2016}, Eq.~\eqref{eq:qfi_hierarchy} enforces a causal speed limit on metrological sensitivity: the local quantum Fisher information is exponentially suppressed outside the Lieb--Robinson light cone. Thus, operator spreading is a necessary condition for distant measurement response. However, as illustrated in Fig.~\ref{fig:schematic}, it is not a sufficient one; an OTOC that has reached its finite-system ceiling~\cite{XuSwingle2019} does not guarantee a large local QFI when number-nonconserving dynamics redistribute the parameter sensitivity into multi-body correlations.

\section{Solvable XX model}\label{sec:analytical_xx}

We now specify the Hamiltonian $\hat H_0$ and study the XX spin chain on $N$ sites with open boundaries, perturbed by a symmetry-breaking on-site field.
\begin{equation}
  \label{eq:hamiltonian}
  \Hhat(h) = J\sum_{i=1}^{N-1} (\hat\sigma_i^x\hat\sigma_{i+1}^x + \hat\sigma_i^y\hat\sigma_{i+1}^y) + h\sum_{i=1}^{N} \hat\sigma_i^x.
\end{equation}
The analysis probes the low-energy single-defect scattering rather than the infinite-temperature or high-density regime typically associated with chaotic scrambling.
Note that a $Z$-field would preserve the $U(1)$ magnon-number symmetry and keep the model free-fermionic; the $X$-field is chosen here to break this symmetry and couple different magnon-number sectors. 
For the polarized initial state and the local $\hat\sigma_y$ encoding used here, the pure XX model ($h=0$) is solvable via a Jordan--Wigner transformation mapping to free fermions~\cite{LiebSchultzMattis1961}, wherein the metrological encoding generates a single propagating quasiparticle. OTOCs in free-fermion chains exhibit ballistic spreading without chaos~\cite{Lin2018,LopezPiqueres2021,Colmenarez2020,XuScaffidi2020}. Operator support spreads across the chain while remaining concentrated in the single-particle sector. Applying the symmetry-breaking field ($h > 0$) generates nonlocal string operators under the Jordan--Wigner map, breaking the free-particle structure and inducing probability leakage across the many-body spectrum. Note that whenever we present numerical results, we use $J=1$.

The Jordan--Wigner transformation maps the pure XX Hamiltonian ($h=0$) to a free-fermion chain, identifying the fully polarized state $\ket{\psi_0} = \ket{\uparrow\cdots\uparrow}$ as the vacuum
denoted by $\ket{\emptyset}$,
\begin{equation}
\Hhat_0 \equiv \Hhat(0) = 2J \sum_{i=1}^{N-1} (\hat c_i^\dagger\hat c_{i+1} + \hat c_{i+1}^\dagger \hat c_i),
\end{equation}
where the factor $2J$ arises from switching to rising/lowering operators, i.e., 
\begin{align}
  \hat\sigma_i^x\hat\sigma_{i+1}^x + \hat\sigma_i^y\hat\sigma_{i+1}^y = 2(\hat\sigma_i^+\hat\sigma_{i+1}^- + \hat\sigma_i^-\hat\sigma_{i+1}^+).
\end{align}
The local metrological encoding $\hat U_s(\theta) = e^{-i\theta \sigs{y}/2}$ perturbs the initial state by generating a single localized spin flip. 
To first order in the parameter $\theta$, this tangent state corresponds to a single fermion injected at the source site
\begin{equation}
\partial_\theta \ket{\psi(\theta)}\big|_{\theta=0} = -\frac{i}{2} \sigs{y} \ket{\emptyset} = \frac{1}{2} \ket{s},
\end{equation}
where the one-magnon basis state at the source is defined as
\begin{equation}
\label{eq:onemagnon}
\ket{s} \equiv c_s^\dagger \ket{\emptyset} = \sigs{-}\ket{\psi_0},
\end{equation}
The pure XX Hamiltonian conserves the total particle number $\sum_i c_i^\dagger c_i$, confining evolution to the single-particle subspace.

The amplitude of the propagation of the localized excitation from the source $s$ to the probe site $j$ after time $t$ is the single-particle Green's function
\begin{equation}\label{eq.green}
  \calG_{j,s}(t) = \bra{\emptyset} c_j e^{-i \Hhat_0 t} c_s^\dagger \ket{\emptyset}.
\end{equation}
The globally evolved encoded state takes the form
\begin{equation}
\ket{\psi_\theta(t)} = \cos\frac{\theta}{2}\ket{\emptyset} + \sin\frac{\theta}{2}\sum_{\ell=1}^N \calG_{\ell,s}(t)\ket{\ell},
\end{equation}
where $\ket{\ell}$ denotes the one-magnon state with the flipped spin at site $\ell$. Tracing out all sites except $j$ yields a $2\times 2$ reduced state whose off-diagonal coherence is proportional to $\calG_{j,s}(t)$. A direct evaluation of the single-site QFI for this one-parameter family at $\theta=0$ gives
\begin{equation}
\label{eq:app_analytical_qfi_base}
F_j(t) = \abs{\calG_{j,s}(t)}^2.
\end{equation}
Although the reduced state is mixed for finite $\theta$, this result is exact at the operating point $\theta=0$ used throughout this work.
The single-particle propagator is unitary, $\sum_{j} \abs{\calG_{j,s}(t)}^2 = 1$, which guarantees the single-site sum rule $\sum_{j} F_j(t) = \Fglob$ at the operating point $\theta=0$. 
At different $\theta$'s the reduced states are mixed with unequal purities and the sum rule does not hold even in the free XX model.

\textit{Infinite chain.}---For an infinite chain, translation invariance dictates that the free-fermion Hamiltonian is diagonalized by the momentum-space Fourier transform
\begin{equation}
 \hat  c_j^\dagger = \int_{-\pi}^\pi\frac{dk}{2\pi}\, e^{-i k j} \hat{\tilde{c}}_k^\dagger,
\end{equation}
where $\hat{\tilde{c}}_k^\dagger$ creates a momentum eigenstate with the single-particle dispersion $E_k = 4J \cos k$. 
Substituting this normal-mode expansion into the Green's function given by Eq.~\eqref{eq.green} allows the time evolution to act diagonally. 
Conservation of lattice momentum collapses the vacuum overlap to a single integral depending only on the relative distance $n=j-s$
\begin{equation}
\calG_{n}(t) = \int_{-\pi}^\pi\frac{dk}{2\pi}\,  e^{i k n} e^{-i 4J t \cos k}.
\end{equation}
The Jacobi--Anger expansion yields
\begin{equation}
  \calG_{n}(t) = (-i)^{n} J_{n}(4Jt),
\end{equation}
where $J_{n}$ is the $n$-th Bessel function of the first order.
The maximal group velocity is $v_{\max}=4J$, underpinning ballistic light cones and thus the speed of local-QFI transport~\cite{Wysocki2025}. 
Substituting the above expression into Eq.~\eqref{eq:app_analytical_qfi_base} gives the ballistic QFI profile $F_j(t) = J_{j-s}^2(4Jt)$.

\textit{Open boundary conditions.}---For a finite open chain of length $N$, the exact standing-wave propagator is
\begin{align}\label{eq:finite_open_prop}
  \calG^{(N)}_{j,s}(t)&= \frac{2}{N+1} \sum_{m=1}^{N} \sin(q_m j)\sin(q_m s)\, e^{-i4Jt\cos q_m},\nonumber\\
  q_m&= \frac{m\pi}{N+1}.
\end{align}
This expression will serve as an $h=0$ analytical benchmark and a free-chain consistency check. 
For $h>0$, the full many-body dynamics are evaluated by exact diagonalization. In the semi-infinite limit ($N \to \infty$ with a wall at $j=0$), the standing-wave propagator reduces to an image-source form,
\begin{equation}
\label{eq:app_bessel_prop}
\calG_{j,s}(t) = (-i)^{j-s} J_{j-s}(4Jt) - (-i)^{j+s} J_{j+s}(4Jt),
\end{equation}
where the first term is the infinite-chain expression while the other one enforces the hard-wall Dirichlet condition at $j=0$. 
The accessible subset of the global QFI under open boundaries therefore scales analytically as
\begin{equation}
\label{eq:app_analytical_qfi_open}
F_j(t) = \abs{J_{j-s}(4Jt) - (-1)^s J_{j+s}(4Jt)}^2.
\end{equation}
For a semi-infinite chain with boundary source $s=1$, this expression simplifies via the Bessel recurrence $J_{n-1}(z)+J_{n+1}(z)=2nJ_n(z)/z$ to
\begin{equation}
\label{eq:qfi_open_s1}
F_j(t)\big|_{s=1} = \frac{j^2}{4J^2 t^2}\,J_j^2(4Jt).
\end{equation}
This analytical baseline is related to the signal-propagation bounds derived for noninteracting tight-binding chains in Ref.~\cite{Wysocki2025}.

The single-particle mapping extends to finite metrological measurement blocks. Writing $A$ for the block and $A^c$ for its complement, define
\begin{equation}
  p_A(t)=\sum_{\ell\in A}|\mathcal G_{\ell,s}(t)|^2 .
\end{equation}
The reduced state on $A$ is
\begin{equation}
  \hat\varrho_A(\theta,t) = \ket{\varphi_A(\theta,t)}\!\bra{\varphi_A(\theta,t)} + \sin^2\frac{\theta}{2}\,[1-p_A(t)]\,\ket{\emptyset_A}\!\bra{\emptyset_A},
\end{equation}
where
\begin{equation}
  \ket{\varphi_A(\theta,t)} = \cos\frac{\theta}{2}\ket{\emptyset_A} + \sin\frac{\theta}{2}\sum_{\ell \in A} \calG_{\ell,s}(t) \ket{\ell}_A.
\end{equation}
The vector $\ket{\varphi_A}$ is intentionally left unnormalized; the second term restores the trace after tracing out the complement.
At $\theta=0$, $\hat\varrho_A = \ket{\emptyset_A}\!\bra{\emptyset_A}$, and the block QFI evaluates to
\begin{equation}
  F_Q(\hat\varrho_A)\big|_{\theta=0} = p_A(t) = \sum_{\ell\in A} \abs{\calG_{\ell,s}(t)}^2 = \sum_{\ell\in A} F_\ell(t).
\end{equation}
Thus, the sensitivity of a finite block is the sum of its single-site components. In this integrable model, the metrological information is encoded in single-particle amplitudes, and it can be concentrated by a local unitary.

\section{Perturbative analysis of $U(1)$ breaking}
\label{sec:perturbation}

In the integrable $U(1)$-symmetric XX model, the local QFI of the encoded probe is carried by a single traveling magnon. Because this continuous symmetry conserves particle number, the initial excitation remains confined within the one-magnon sector, preventing leakage into higher-magnon sectors. 

Introducing the symmetry-breaking field $h>0$ breaks the block-diagonal structure of the magnon-number sectors, coupling the one-magnon tangent state to both the vacuum and two-magnon sectors. The isolated single-particle component leaks into multi-magnon states, redistributing the sensitivity. By treating this coupling perturbatively, we will show that the leakage is constrained to $\mathcal{O}(h^2)$ by sector orthogonality and the spectral structure of the reduced probe state, producing the temporal separation of $F_k$, $F_\mathrm{dec}$, and $F_Q(\hat\varrho_{\mathcal{N}_w})$ observed in the variational decoder numerics.
(For the definition of the block size $\mathcal N_w$, see the caption of Fig.~\ref{fig:schematic}). 

\textit{Vanishing of the first-order correction.}---We calculate the QFI from the reduced single-qubit density matrix at the target probe site $k$. The global initial state is prepared by a weak generator rotation at the source site $s$, giving the initial pure state
\begin{equation}
\ket{\psi(\theta)} = e^{-i \theta \hat\sigma_y^{(s)}/2} \ket{\psi_0} \approx \ket{\psi_0} + \frac{\theta}{2} \ket{s},
\end{equation}
where we define the encoded one-magnon tangent state as $\ket{s} = -i\hat\sigma_y^{(s)}\ket{\psi_0}$, which coincides with $\hat\sigma_s^-\ket{\psi_0}$ for the polarized reference state (here magnons are down-spins relative to the fully $z$-polarized up state, so $\hat\sigma^- \ket{\uparrow}=\ket{\downarrow}$). At the reference value $\theta=0$, the unperturbed initial state is strictly the vacuum $\ket{\psi_0}$. The leading metrological response at a distant site $k$ is therefore determined by the first derivative of the local Bloch vector $\mathbf{r}_k(t, \theta)$. The expectation value of the local Pauli observables $r_k^\alpha = \bra{\psi(\theta)} \hat\sigma_\alpha^{(k)}(t) \ket{\psi(\theta)}$ differentiated 
with respect to $\theta$ gives the linear tangent overlap
\begin{equation}
\label{eq:bloch_deriv_exact}
\partial_\theta r_k^\alpha \big|_{\theta=0} = \mathrm{Re}\langle\psi_0|\hat\sigma_\alpha^{(k)}(t)|s\rangle,
\end{equation}
where $\hat\sigma_\alpha^{(k)}(t) = e^{i\Hhat t}\hat\sigma_\alpha^{(k)} e^{-i\Hhat t}$ is the fully interacting, Heisenberg-evolved probe observable driven by the perturbed Hamiltonian $\Hhat = \Hhat_0 + h\hat V$. 

To isolate the effect of the $U(1)$-breaking perturbation, we move to the interaction picture with respect to the integrable baseline $\Hhat_0$. The full observable expands formally as
\begin{equation}
  \hat\sigma_\alpha^{(k)}(t) = \hat{\mathcal{U}}_I^\dagger(t) \, \hat\sigma_\alpha^{(k)}(t)_0 \, \hat{\mathcal{U}}_I(t),
\end{equation}
where $\hat\sigma_\alpha^{(k)}(t)_0 = e^{i\Hhat_0 t}\hat\sigma_\alpha^{(k)} e^{-i\Hhat_0 t}$ is the trivially solvable free-fermion probe evolution. Expanding the associated interaction-picture propagator 
\begin{align}
  \hat{\mathcal{U}}_I(t) = \mathcal{T} \exp\left(-i h \int_0^t \hat V_I(t') dt'\right)
\end{align}
to first order in $h$ gives the leading commutator correction
\begin{equation}
\label{eq:dyson_obs}
\hat\sigma_\alpha^{(k)}(t) = \hat\sigma_\alpha^{(k)}(t)_0 + ih \int_0^t dt_1\, [\hat V_I(t_1), \hat\sigma_\alpha^{(k)}(t)_0] + \mathcal{O}(h^2).
\end{equation}

In the integrable $h=0$ limit, we evaluate the exact baseline sensitivity via Eq.~\eqref{eq:bloch_deriv_exact} by decomposing the action of the probe observable on the one-magnon basis. The transversal spin operator connects the single magnon to the vacuum and the two-magnon sectors
\begin{equation}
\hat\sigma_y^{(k)}\ket{j} = -i\delta_{jk}\ket{\emptyset} + i(1-\delta_{jk})\ket{j,k}.
\end{equation}
When we project this action onto the vacuum bra $\bra{\psi_0}$ required by the Bloch derivative, the two-magnon term drops out. 
The inner product reduces to the single-particle Green's function $\calG_{k,s}(t) = \bra{k} e^{-i\Hhat_0t} \ket{s}$
\begin{equation}
\bra{\psi_0} \hat\sigma_y^{(k)}(t)_0 \ket{s} = \langle\emptyset|e^{i\Hhat_0t} \hat\sigma_y^{(k)} e^{-i\Hhat_0t}|s\rangle = -i \calG_{k,s}(t).
\end{equation}
Taking the real part gives the leading-order Bloch vector derivatives
\begin{align}
\partial_\theta r_k^y \big|_{\theta=0} &= \mathrm{Im}\,\calG_{k,s}(t), \\
\partial_\theta r_k^x \big|_{\theta=0} &= \mathrm{Re}\,\calG_{k,s}(t).
\end{align}
Summing these quadrature components yields the tangent norm $|\partial_\theta \mathbf{r}_k|^2 = |\calG_{k,s}(t)|^2$. Because the unperturbed local state is the vacuum, the mixedness from tracing out the complement contributes only at $\mathcal{O}(\theta^2)$. The zeroth-order QFI therefore equals the wavepacket density
\begin{equation}
F_k^{(0)}(t) = |\calG_{k,s}(t)|^2.
\end{equation}

For the transverse components $\alpha = x,y$, the first-order $\mathcal{O}(h)$ correction to the Bloch-vector derivative vanishes
\begin{equation}
\label{eq:first_order_interf}
\partial_\theta r_k^\alpha \big|^{(1)} = \mathrm{Re} \left( i h \int_0^t dt_1\, \bra{\psi_0} [\hat V_I(t_1), \hat\sigma_\alpha^{(k)}(t)_0] \ket{s} \right) = 0
\end{equation}
This cancellation follows from the $U(1)$ particle-number conservation of the free XX model. By denoting $\mathcal{H}^{(n)}$ the $n$-magnon subspace, 
the initial states are $\bra{\psi_0} \in (\mathcal{H}^{(0)})^\dagger$ and $\ket{s} \in \mathcal{H}^{(1)}$, and the free propagator $U_{\mathrm{XX}}$ preserves these particle sectors.

For the first term in the commutator, $\bra{\psi_0}\hat V_I(t_1) \hat\sigma_\alpha^{(k)}(t)_0 \ket{s}$, 
the transversal probe operators $\hat\sigma_{x,y}^{(k)}$ flip a single spin, coupling the one-magnon ket down to the zero-magnon sector or up to the two-magnon sector: 
$\hat\sigma_\alpha^{(k)}(t)_0 \ket{s} \in \mathcal{H}^{(0)} \oplus \mathcal{H}^{(2)}$. Meanwhile, the perturbation $\hat V_I(t_1) = \sum_j \hat\sigma_j^x(t_1)_0$ 
acting on the vacuum bra creates a single magnon, producing $\bra{\psi_0}\hat V_I(t_1) \in (\mathcal{H}^{(1)})^\dagger$. 
The overlap between $\mathcal{H}^{(1)}$ and $\mathcal{H}^{(0)} \oplus \mathcal{H}^{(2)}$ is strictly zero.
For the second term, $\bra{\psi_0} \hat\sigma_\alpha^{(k)}(t)_0 V_I(t_1) \ket{s}$, the logic is inverted. The probe operator acting on the vacuum bra yields $\bra{\psi_0}\hat\sigma_\alpha^{(k)}(t)_0 \in (\mathcal{H}^{(1)})^\dagger$, while the perturbation acting on the one-magnon ket yields $V_I(t_1)_0\ket{s} \in \mathcal{H}^{(0)} \oplus \mathcal{H}^{(2)}$. States from different sectors are orthogonal, thus
the transverse contribution to the tangent norm has no linear correction in $h$.

The sector-orthogonality argument above applies directly to $\alpha \in \{x,y\}$. For $\alpha = z$, the zeroth-order derivative vanishes, $\partial_\theta r_k^z|_{\theta=0}^{(0)} = 0$, because a number-conserving operator cannot connect the zero- and one-magnon sectors. While the $\mathcal{O}(h)$ correction to $r_k^z$ is generically nonzero, it enters the tangent norm quadratically. Thus, the tangent norm has no linear correction in $h$:
\begin{equation}
\label{eq:tangent_norm_h2}
|\partial_\theta \mathbf{r}_k|^2 = |\calG_{k,s}(t)|^2 + \mathcal{O}(h^2).
\end{equation}

\textit{Control of the radial Bloch-vector contribution.}---The full single-qubit QFI
contains a radial correction $(\mathbf{r}_k \cdot \partial_\theta \mathbf{r}_k)^2/(1-|\mathbf{r}_k|^2)$
whose denominator vanishes at the pure-state point $h=0$~\cite{Safranek2017}.
To control this term, we use the spectral representation of the QFI.
For a qubit with eigenvalues $\lambda_\pm$ and eigenstates $\ket{\pm}$,
\begin{equation}
F_Q(\varrho_k) = 4|D_{+-}|^2 + \frac{(\partial_\theta \lambda_-)^2}{\lambda_-(1-\lambda_-)},
\end{equation}
where $D_{+-} = \bra{+}\partial_\theta\hat\varrho_k\ket{-}$ is the off-diagonal matrix element
of the tangent and all matrix elements are evaluated in the eigenbasis of $\hat \varrho_k(\theta=0,h)$.
The first term is the coherent (transverse) contribution;
the second is the potentially singular radial contribution.

To determine the scaling of $\lambda_-$, we expand the evolved reference state at $\theta=0$. Writing $\ket{\Omega_h(t)} = e^{-i\Hhat(h)t}\ket{\psi_0}$, the Dyson expansion gives
\begin{equation}
\ket{\Omega_h(t)} = \ket{\emptyset} + h\sum_\ell A_\ell(t)\ket{\ell} + \mathcal{O}(h^2),
\end{equation}
where the first-order correction lies entirely in the one-magnon sector because $\hat V\ket{\emptyset} = \sum_i\ket{i} \in \mathcal{H}^{(1)}$. The reduced single-site state at $\theta=0$ therefore 
has the form $\hat\varrho_k = \ket{\!\uparrow}\!\bra{\uparrow} + \mathcal{O}(h)$, with off-diagonal elements at $\mathcal{O}(h)$ and diagonal corrections at $\mathcal{O}(h^2)$. The small eigenvalue of a $2\times 2$ matrix near a rank-1 projector is controlled by the determinant:
\begin{equation}
  \lambda_- = \det\left[\hat \varrho_k\right] + \mathcal{O}((\det\left[\hat \varrho_k\right])^2).
\end{equation}
For a global state of the form $\ket{\Omega_h} \approx A_0\ket{\emptyset} + \sum_\ell a_\ell \ket{\ell}$ with one-magnon amplitudes $a_\ell = hA_\ell$, the reduced-state determinant at site $k$ evaluates to
\begin{equation}
  \det\left[\hat \varrho_k\right]= |a_k|^2 \sum_{\ell \neq k}|a_\ell|^2 + \mathcal{O}(|a|^6).
\end{equation}
Since $a_\ell = \mathcal{O}(h)$, this gives $\lambda_- = \mathcal{O}(h^4)$.

The $\theta$-derivative of $\lambda_-$ inherits the same product structure.
Because the encoding shifts the one-magnon amplitudes as
$a_\ell \to hA_\ell + (\theta/2)\calG_{\ell,s}(t) + \mathcal{O}(h\theta)$,
differentiating the determinant with respect to $\theta$ at $\theta=0$
produces terms of the form $h \cdot h^2$ from the product rule,
yielding $\partial_\theta\lambda_- = \mathcal{O}(h^3)$.
The first-order tangent also contains a vacuum component
and a two-magnon component at $\mathcal{O}(h\theta)$.
The vacuum component modifies only the large-eigenvalue branch
of the reduced probe state to this order
and does not create local mixedness.
The two-magnon component has zero overlap
with the vacuum reference under the one-site partial trace,
so it gives no $\mathcal{O}(h)$ contribution.
Its overlap with the $\mathcal{O}(h)$ one-magnon correction
to the reference state can enter the reduced density matrix
only at $\mathcal{O}(h^2)$.
In the local eigenbasis near the rank-one projector,
these terms are off-diagonal with respect to
the small-eigenvalue direction at leading order,
so their contribution to $\partial_\theta\lambda_-$
is at most $\mathcal{O}(h^3)$.
Under the same sector-orthogonality assumptions, the first nonzero contribution to $\partial_\theta\lambda_-$ arises at order $h^3$.
Consequently, the radial QFI contribution scales as
\begin{equation}
\frac{(\partial_\theta\lambda_-)^2}{\lambda_-(1-\lambda_-)}
= \frac{\mathcal{O}(h^6)}{\mathcal{O}(h^4)} = \mathcal{O}(h^2).
\end{equation}

For the coherent term, the transverse tangent norm has no linear correction by the sector argument above. Equivalently,
\begin{equation}
4|D_{+-}|^2 = |\mathcal G_{k,s}(t)|^2+\mathcal O(h^2),
\end{equation}
up to the $\mathcal O(h)$ rotation of the local eigenbasis.
Combining both contributions gives
\begin{equation}
\label{eq:qfi_h2}
F_k(t;h) = |\calG_{k,s}(t)|^2 + \mathcal{O}(h^2)
\end{equation}
for fixed $t$, or more generally within the pre-secular perturbative regime
$t \ll J/h^2$.

\textit{Scattering into the two-magnon continuum.}---At the reference parameter $\theta=0$, the local encoding $\hat U_s(\theta) = e^{-i\theta \sigs{y}/2}$ generates the isolated tangent state
\begin{equation}
\ket{\chi(0)} = \partial_\theta \hat U_s(\theta)\ket{\psi_0}\big|_{\theta=0} = \frac{1}{2}\ket{s},
\end{equation}
where $\ket{s} = \hat\sigma_s^-\ket{\psi_0}$. The evolved tangent state is obtained by applying the full system evolution. Working in the interaction picture with respect to $\Hhat_0$, 
we define the interaction-picture perturbation $\hat V_I(t') = e^{i\Hhat_0t'} V e^{-i\Hhat_0t'}$, where $\hat V = \sum_i \hat\sigma_i^x$. 
Expanding the corresponding unitary sequence to first order in the amplitude $h$ gives
\begin{align}
  \ket{\chi(t)}&= \frac{1}{2}\ket{\phi_1^{(0)}(t)}\\
  &-\frac{i h}2 \int_0^t dt' \, e^{-i\Hhat_0(t-t')}\Big(\sum_i \hat\sigma_i^x\Big)\ket{\phi_1^{(0)}(t')}+ \mathcal{O}(h^2).\nonumber
\end{align}
Here, the zeroth-order output piece $\ket{\phi_1^{(0)}(t)} = e^{-i\Hhat_0t}\ket{s}$ defines the free one-magnon wavepacket
\begin{equation}
\ket{\phi_1^{(0)}(t)} = \sum_{\ell} \mathcal{G}_{\ell,s}(t)\ket{\ell}.
\end{equation}
Applying the transverse field $\sum_i \hat\sigma_i^x = \sum_i (\hat\sigma_i^+ + \hat\sigma_i^-)$ on this single-magnon state splits the first-order correction into two distinct sectors. The $\hat\sigma^+$ ladder operators annihilate the magnon to generate a vacuum interference channel. The vacuum-channel contribution to the first-order correction is
\begin{equation}
\ket{\delta\phi_0(t)} = -ih \int_0^t dt'\, \sum_\ell \mathcal{G}_{\ell,s}(t')\ket{\emptyset}.
\end{equation}
For the infinite-chain propagator, $\sum_\ell \mathcal{G}_{\ell,s}(t') = e^{-i4Jt'}$, giving
\begin{equation}
\|\delta\phi_0(t)\|^2 = \frac{h^2}{8J^2}(1-\cos 4Jt).
\end{equation}
This displays bounded oscillations, contributing no macroscopic continuous decay. (For finite open chains this identity is modified by boundary effects; the expression is used here only to illustrate that the vacuum channel is oscillatory and nonsecular.)
The $\hat\sigma^-$ creation operators spawn a secondary magnon at each site $i \ne \ell$, generating a two-magnon channel $\ket{\delta\phi_2(t)}$. The two-magnon amplitude disperses, and the leakage norm involves nonlocal two-body overlaps
\begin{align}
  \|\delta\phi_2(t)\|^2&= h^2 \iint_0^t dt_1dt_2\\
  &\sum_{i\neq\ell, i'\neq\ell'}\mathcal{G}_{\ell,s}^*(t_1)\mathcal{G}_{\ell',s}(t_2)\langle i,\ell|e^{i\Hhat_0(t_1-t_2)}|i',\ell'\rangle\nonumber.
\end{align}
For short times, the leakage grows quadratically in time before crossing over to a secular golden-rule-like regime. As derived in Appendix~\ref{app:block_magnon}, this gives way to an intermediate regime where the discrete sum approximates a continuous integral weighted by the two-magnon density of states, yielding a golden-rule-like perturbative scaling
\begin{equation}
\label{eq:fgr_decay_main}
\|\delta\phi_2(t)\|^2 \sim \Gamma t \propto \frac{h^2}{J} t.
\end{equation}
In the finite chains studied numerically this should be interpreted as a perturbative scaling estimate rather than an asymptotic decay law.

\textit{Second-order decay of the local QFI.}---Because the transverse tangent contribution has no linear correction and the vacuum channel remains bounded, the leading sustained suppression of the local metrological signal comes from leakage into the two-magnon sector. At first order, the tangent vector decomposes as
\begin{equation}
\ket{\chi(t)} \approx \frac{1}{2}\left[\ket{\phi_1^{(0)}(t)} + \ket{\delta\phi_0(t)} + \ket{\delta\phi_2(t)}\right] + \mathcal{O}(h^2).
\end{equation}
Tracing out the complement reduces the portion of $\partial_\theta \hat\varrho_j$ that remains coherent with the original one-magnon channel. 
This two-magnon leakage drives a second-order depletion of the locally recoverable single-particle contribution.
At the level of the single-particle survival weight, this suggests a depletion
\begin{equation}
P_1(t)\simeq 1-\mathcal O\!\left(\frac{h^2t}{J}\right)
\end{equation}
within the perturbative time window. Consequently, the local QFI profile is expected to show an $\mathcal O(h^2t/J)$ suppression of its one-magnon contribution, although the detailed site-resolved profile need not be given by a purely multiplicative survival factor. This is consistent with the leading-order expansion of a survival probability $P(t) \approx e^{-\Gamma t}$ with $\Gamma \propto h^2/J$. Beyond the characteristic scattering timescale $t \gtrsim J/h^2$, the single-particle description breaks down and the sensitivity is redistributed into multiparticle correlations.

\begin{figure}[t!]
\centering
\includegraphics[width=\columnwidth]{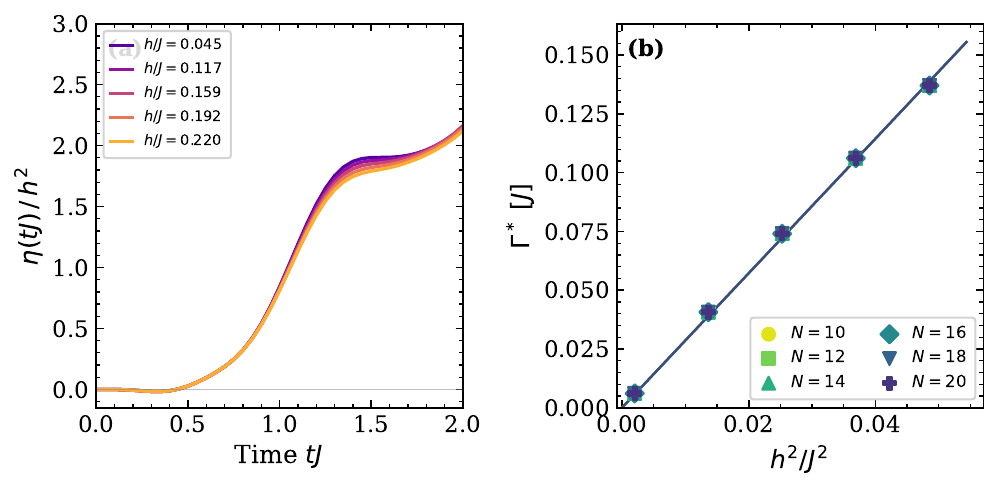}
\caption{(a)~Rescaled depletion $\eta(t)/h^2$ for several perturbation strengths at $N=20$; the collapse is consistent with $\eta \propto h^2$ from the two-magnon perturbative analysis. (b)~Growth rate $\Gamma^*$ versus $h^2/J^2$ for $N = 10$--$20$, converging to $\Gamma^* \approx 2.86\,h^2/J$.}
\label{fig:fgr_ratio}
\end{figure}

The second-order depletion $1 - \mathcal{O}(h^2 t/J)$ can be probed through the site-summed local-QFI ratio
\begin{equation}\label{eq:eta_def}
\eta(t) \;=\; 1 \;-\; \frac{\sum_j F_j(t;h)}{\sum_j F_j(t;0)}.
\end{equation}
The equality $\sum_j F_j(t;0) = F_\mathrm{glob}$ is specific to the $\theta=0$ free XX benchmark; for $h>0$, the site-summed QFI is a diagnostic that need not equal a conserved probability. If the leading depletion is controlled by the two-magnon leakage norm, the rescaled depletion $\eta/h^2$ should collapse to a single function of time in the perturbative window. Figure~\ref{fig:fgr_ratio}(a) verifies this collapse across $h/J \in [0.045, 0.220]$ at $N=20$. Extracting the instantaneous growth rate $\Gamma^* = d\eta/dt$ over $0.8 \le tJ \le 1.2$ and plotting it against $h^2/J^2$ for chains of length $N = 10$--$20$ [Fig.~\ref{fig:fgr_ratio}(b)] yields a proportionality $\Gamma^* \approx 2.86\,h^2/J$, converged for $N \geq 12$. We use $\Gamma^*$ as an empirical perturbative rate characterizing early-time leakage, not as an asymptotic thermodynamic decay constant.

\section{Variational block decoder}\label{sec:decoder}

To quantify how much metrological information is recoverable by a specified local decoder, we consider a parameterized unitary $\hat V_\phi$ acting on the $w$-body reduced density matrix 
$\hat\varrho_{\mathcal{N}_w}$ of a contiguous neighborhood. Because any unitary plus partial trace is a quantum channel, the sensitivity extracted onto the 
final probe site $k$ is bounded by the exact block capacity. 
\begin{figure*}[!t]
\centering
\includegraphics[width=\textwidth]{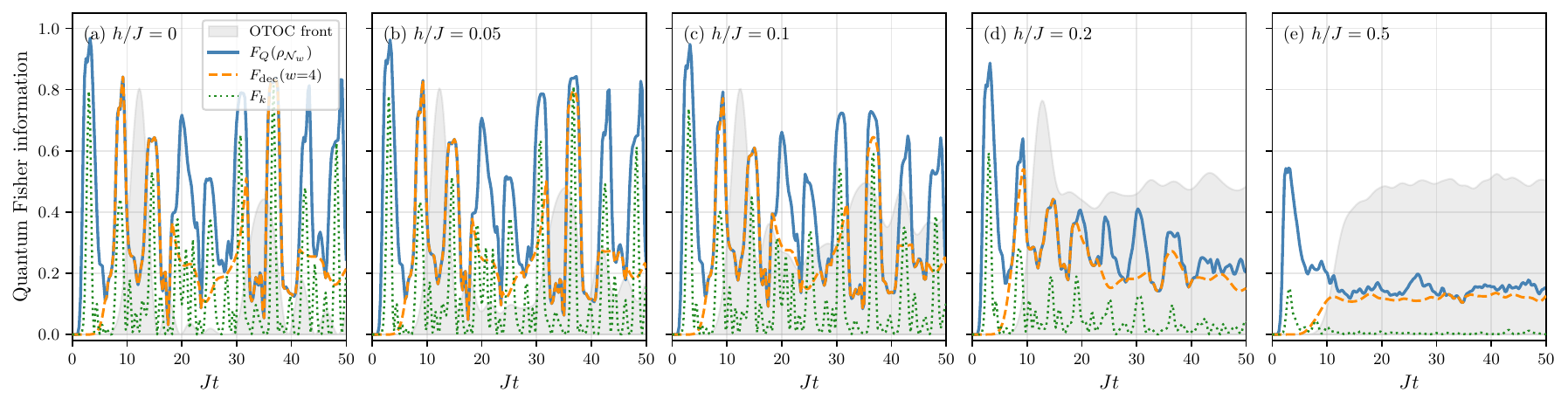}
\caption{Nested QFI hierarchy at probe site $k=10$ for an $N=10$ chain with $S^y$ encoding at $s=1$. The measurement block $\mathcal{N}_w$ covers sites $\{7,8,9,10\}$. Blue (solid): exact block 
  QFI $F_Q(\hat\varrho_{\mathcal{N}_w})$. Orange (dashed): decoded QFI $F_\mathrm{dec}$ optimized over the sequential sweep decoder. Green (dotted): bare single-site QFI $F_k$. The shaded gray region is the squared commutator $C_{sk}^{(y)}$ (scaled by $1/2$), marking the arrival of the operator lightcone at the probe. Panel~(a): integrable limit ($h=0$). Panels~(b)--(e): increasing $h/J$, illustrating the suppression of $F_\mathrm{dec}$ and $F_Q(\hat\varrho_{\mathcal{N}_w})$ relative to $F_\mathrm{glob}=1$.}
\label{fig:blockqfi}
\end{figure*}
The decoded QFI is defined by the maximum over $\phi$.
Numerically, we report the best value found by the optimization
procedure described below, which is therefore a lower bound
on the formal optimum.
\begin{equation}
\begin{split}
  F_k(t) &\le \max_\phi F_Q\Big(\Tr_{\calN_w \setminus \{k\}} \hat V_\phi \hat\varrho_{\calN_w} \hat V_\phi^\dagger \Big) \\
  &\equiv F_{\mathrm{dec}}(w,t) \le F_Q(\hat\varrho_{\mathcal{N}_w}) \le F_\mathrm{glob}.
\end{split}
\end{equation}
The gap between the bare probe sensitivity $F_k$ and the decoded sensitivity $\Fdec$ measures information recovered from local block correlations. The margin between $\Fdec$ and the exact block QFI $F_Q$ has a fundamental information-theoretic component: applying a unitary decoder followed by a partial trace maps the block state to a two-dimensional output system. Therefore only the component of the block's parameter dependence that can be represented on an effective qubit can be preserved. At $h=0$, the local parameterized block-state family has support only on the vacuum and a single one-magnon direction, i.e.\ on an effective two-dimensional subspace. A single output qubit therefore suffices to capture the full block QFI. At $h>0$, leakage into multi-magnon sectors generically increases the rank of the reduced block state and enlarges the support of the SLD beyond an effectively two-dimensional subspace. In that case, a unitary followed by single-qubit readout is generically unable to preserve the full block QFI, because the relevant SLD support need not be compressible into a two-dimensional output space. Thus the observed gap reflects a fundamental information-theoretic dimension bottleneck, in addition to any residual ansatz or optimization limitations.

We implement $\hat V_\phi$ as a sequential ``sweep'' circuit of general $SU(4)$ two-qubit gates moving directionally toward the target
\begin{equation}
\label{eq:sweep_circuit}
\hat V_\phi = \hat U^{(k-1, k)}(\phi_{w-1}) \cdots \hat U^{(k-w+1, k-w+2)}(\phi_1).
\end{equation}
This $\mathcal{O}(w)$-depth sequence acts as a directional single-magnon filter. As established in Sec.~\ref{sec:analytical_xx}, at $h=0$ the reduced block-state family has support on the vacuum and one collective one-magnon direction. Because any single-excitation superposition (analogous to a W-state) can be perfectly compressed onto a single qubit using a linear sequence of nearest-neighbor two-qubit gates~\cite{Bartschi2019, Yang2025}, the sweep circuit is kinematically exact for routing arbitrary one-magnon states. It therefore achieves fully saturated recovery $F_\mathrm{dec} = F_Q(\hat\varrho_{\mathcal{N}_w})$ in the integrable limit to numerical precision, validating both the ansatz and the optimization pipeline.

Each two-qubit gate is parameterized as a general $SU(4)$ unitary,
generated by a traceless anti-Hermitian $4\times 4$ matrix
with 15 independent real parameters.
The exact block QFI is computed from the eigendecomposition
of the reduced density matrix using the spectral formula
\begin{align}
  F_Q(\hat\varrho) = 2\!\!\!\!\!\sum_{\substack{\mu,\nu:\\ \lambda_\mu+\lambda_\nu>0}}\!\!\!\!\!
  \frac{|\langle\mu|\partial_\theta\rho|\nu\rangle|^2}{\lambda_\mu+\lambda_\nu},
\end{align}
where $\ket{\nu}$ denote the eigenvectors with the corresponding eigenvalues $\lambda_\nu$.
The derivative $\partial_\theta\hat\varrho_{\calN_w}$ is evaluated
analytically from the tangent state at $\theta=0$.
The circuit parameters are optimized via automatic differentiation
(Adam, 300 steps).
The $h=0$ saturation $F_\mathrm{dec} = F_Q$ validates convergence
in the one-magnon sector;
the remaining gap at $h>0$ reflects generic finite-dimensional
compression limitations and possible residual optimization error.

\begin{figure*}[!t]
\centering
\includegraphics[width=0.8\linewidth]{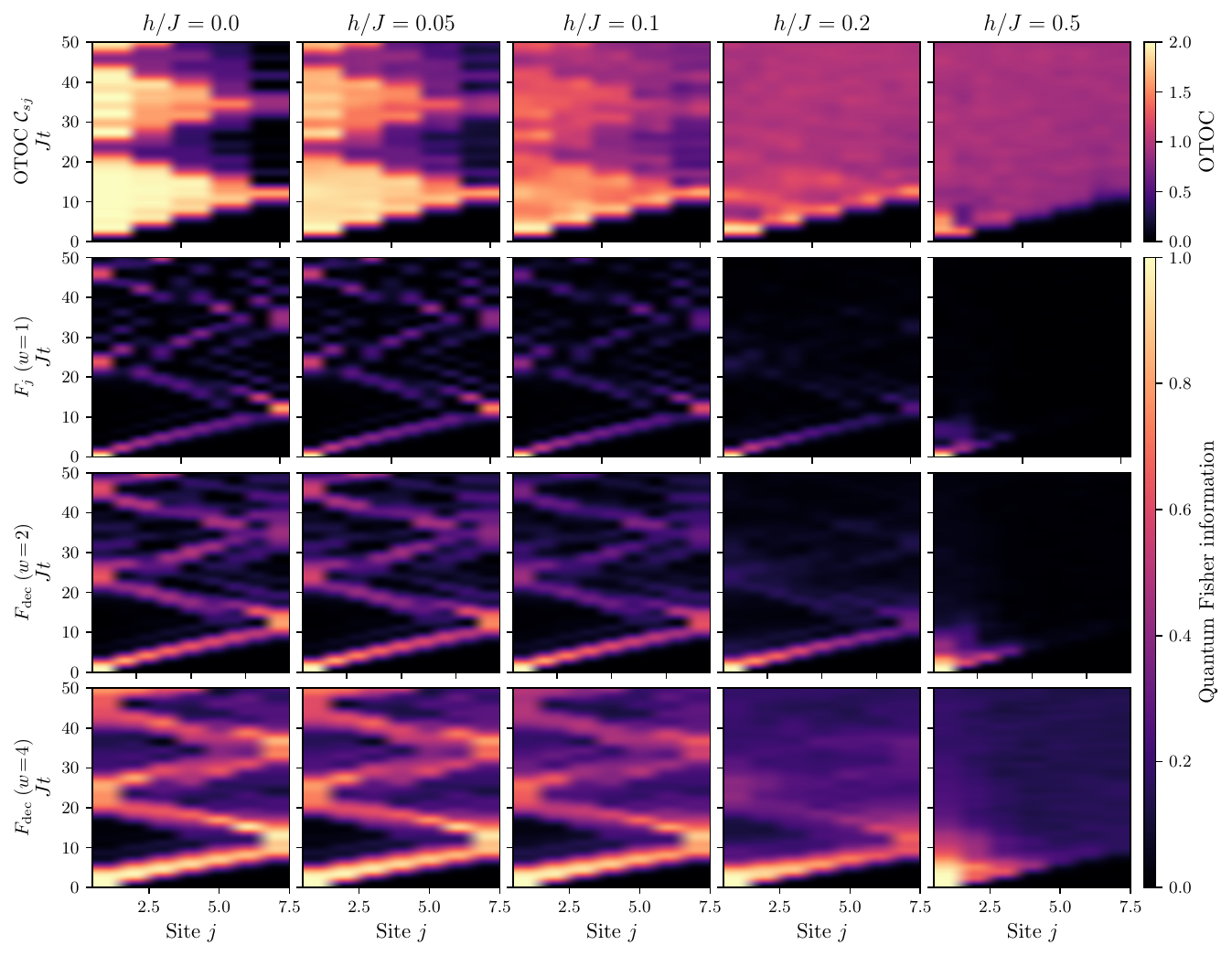}
\caption{Spatiotemporal diagnostics of metrological sensitivity
and its local recoverability for an $N=10$ chain with $S^y$ encoding.
Columns correspond to increasing $U(1)$-breaking field strength
$h/J \in \{0,\, 0.05,\, 0.1,\, 0.2,\, 0.5\}$.
\textbf{Row~1:} Representative squared-commutator component
$C_{sj}^{(y)}(t)$ (colorbar $0$--$2$),
shown as a diagnostic of the operator front
rather than as the rigorous summed bound $\calC_{sj}^{\mathrm{sum}}$.
\textbf{Row~2:} Raw single-site QFI $F_j(t)$.
\textbf{Row~3:} Decoded single-qubit QFI $\Fdec(w\!=\!2,t)$
after a two-site block decoder.
\textbf{Row~4:} Decoded QFI $\Fdec(w\!=\!4,t)$
after a four-site block decoder.
In the integrable limit ($h=0$) the QFI traces the OTOC front.
As $h$ increases, the raw single-site QFI is suppressed
while the plotted squared-commutator component
reaches its finite-system ceiling.
The block decoders partially recover the sensitivity at weak $h$,
but remain suppressed at strong $U(1)$ breaking ($h=0.5$),
consistent with multi-magnon or collective structure
not extracted by the sweep decoder.}
\label{fig:results}
\end{figure*}

Figure~\ref{fig:results} compares the OTOC, the raw single-site QFI, and the decoded QFI for block widths $w=2$ and $w=4$ across five values of the $U(1)$-breaking field $h$. At $h=0$, the decoded QFI with $w=4$ tracks the same ballistic front as the OTOC, consistent with the exact result of Sec.~\ref{sec:analytical_xx}: $F_{\mathrm{dec}}(w,t) = F_Q(\hat\varrho_{\mathcal{N}_w}(t))$ and the decoder recovers all block sensitivity onto the output qubit.

For weak fields ($h/J \lesssim 0.1$), the raw single-site QFI no longer follows the squared-commutator front, but the $w=4$ decoder recovers a large fraction of the exact block QFI. This is consistent with the perturbative analysis of Sec.~\ref{sec:perturbation}: the leading $\mathcal{O}(h^2)$ leakage populates two-magnon states that remain within the spatial extent of the block, so the decoder recovers a substantial fraction of the block sensitivity. The $w=2$ decoder captures less of the signal,
consistent with sensitivity being distributed over correlations
extending beyond two sites.

At $h/J = 0.2$, the decoded QFI begins to separate from the squared-commutator footprint at late times---the reduced $w=4$ block no longer contains or compresses the relevant sensitivity as effectively. At $h/J = 0.5$, the plotted squared-commutator component $C_{sj}^{(y)}$ reaches its finite-system ceiling across the chain while the decoded QFI is uniformly suppressed. The perturbative regime has broken down: the sensitivity is distributed across the system and the sweep decoder, supported on a block of width $w\le 4$, cannot recover a significant fraction of it over the studied time window.

Figure~\ref{fig:blockqfi} provides a complementary view of the same hierarchy, plotting the three quantities $F_k$, $F_\mathrm{dec}(w=4)$, and $F_Q(\hat\varrho_{\mathcal{N}_w})$ 
simultaneously as functions of time at the fixed probe $k = N$ (opposite boundary from the encoding site $s=1$). The block $\mathcal{N}_w$ covers sites $[N-3,\ldots,N]$. Because the block extends spatially toward the encoding site, the ballistic front reaches the proximal block edge before the terminal boundary site, and the block QFI activates correspondingly earlier. In the integrable limit [panel~(a)], the orange decoded QFI tracks the blue block QFI closely, confirming that the sequential sweep circuit is a near-exact concentrator in the one-magnon sector. As $h$ increases, the gap between $F_\mathrm{dec}$ and $F_Q$ widens progressively. Since the sweep decoder is kinematically exact in the one-magnon sector, this gap reflects the rank-based compression limit discussed above: the block state at $h>0$ has rank greater than~2, and a single output qubit cannot capture the full block QFI. At $h/J=0.5$ [panel~(e)], the exact block QFI $F_Q(\rho_{\mathcal{N}_w})$ is sharply suppressed relative to $F_\mathrm{glob}=1$. Since $F_Q(\rho_{\mathcal{N}_w})$ bounds any operation on the block, the reduced state of the chosen neighborhood no longer contains the full distinguishability; the missing QFI resides in the complement or in correlations with it.

The decoder calculations are performed at $N=10$, where exact diagonalization and exact reduced-state QFI are feasible for all time points. For the finite open chain, the exact standing-wave propagator of Eq.~\eqref{eq:finite_open_prop} incorporates both boundaries at all times; the hierarchy and the decoder gaps are well defined throughout. While dynamics at $tJ \gtrsim 2.5$ feature finite-size boundary reflections, the hierarchy of recoverability remains perfectly valid as an operational metric regardless of these boundary effects. The perturbative depletion scaling is checked up to $N=20$. We interpret the decoder results as finite-size operational diagnostics of recoverability, not as an asymptotic thermodynamic characterization.

\section{Discussion and conclusions}\label{sec:discussion}

While out-of-time-order correlators diagnose the spreading of operator support and establish an effective Lieb--Robinson light cone outside which quantum Fisher information is exponentially suppressed, they do not determine whether parameter sensitivity within that light cone remains locally recoverable.
In this work, we have demonstrated that operator spreading and metrological sensitivity are intertwined and that their distinction is determined by the breakdown of magnon-number conservation.
By comparing the bare probe QFI, the variational QFI and the exact block QFI, we have established a quantitative hierarchy of local metrological accessibility.
In the integrable free-fermion limit, ballistic spreading preserves local recoverability; a shallow unitary decoder can concentrate the extended block sensitivity onto a single output qubit.
However, the $U(1)$--breaking transverse field scatters the initial excitation into multi-magnon sectors, driving a progressive gap between the exact block QFI and the decoder output.

The hierarchy used here is independent of the particular
XX-chain realization.
For any locally encoded parameter followed by
parameter-independent unitary dynamics,
the conserved global QFI can be decomposed operationally
into bare local sensitivity, decoder-recoverable sensitivity,
finite-block sensitivity, and globally stored sensitivity.
The gaps between these quantities have distinct meanings:
$F_{\mathrm{dec}} - F_k$ measures sensitivity that can be recovered
from nearby correlations,
$F_Q(\hat\varrho_{\mathcal{N}_w}) - F_{\mathrm{dec}}$ measures
sensitivity that is present in the block but not extracted
by the chosen decoder,
and $F_{\mathrm{glob}} - F_Q(\rho_{\mathcal{N}_w})$ measures
sensitivity unavailable to operations supported on that block.
In the XX chain these gaps vanish within the one-magnon sector,
while the transverse field opens them by coupling the tangent state
to higher-magnon sectors.

This approach complements recent studies tracking the quantum Fisher information (QFI) of subsystems in noninteracting chains~\cite{FerroFagotti2025}
and recovery protocols in scrambling systems~\cite{YoshidaYao2019,HaydenPreskill2007}.
Using a sequential decoder that is kinematically exact in the one-magnon sector allows us to isolate the onset of sensitivity redistribution driven specifically by $U(1)$--breaking dynamics.
This provides a more refined metrological diagnostic of operator spreading than the squared commutator alone, as it distinguishes the arrival of operator support 
from the local recoverability of the parameter sensitivity carried by that support.

\begin{acknowledgments}
M.P.\ acknowledges funding from MICIU/AEI/10.13039/501100011033/FEDER, UE.
\end{acknowledgments}

% ═══════════════════════════════════════════════════════════════
%  APPENDICES
% ═══════════════════════════════════════════════════════════════
\appendix

\section{Single-qubit quantum Fisher information}\label{app:bloch}

For a single-qubit state parameterized by the Bloch vector $\mathbf{r}(\theta)$, the density matrix is 
$\hat\varrho(\theta) = \frac{1}{2}(\mathbb{I} + \mathbf{r}(\theta) \cdot \boldsymbol{\hat\sigma})$. The symmetric logarithmic derivative (SLD) $\hat L_{\theta}$ satisfies 
$\partial_\theta\hat\varrho = \frac{1}{2}(\hat L_\theta\hat\varrho+\hat\varrho\hat L_\theta)$. In the Bloch representation, the SLD can be expressed directly in terms of the Bloch vector and its derivative
\begin{equation}
  \hat L_\theta = (\partial_\theta \mathbf{r}) \cdot \boldsymbol{\hat\sigma} + \frac{\mathbf{r} \cdot \partial_\theta \mathbf{r}}{1 - |\mathbf{r}|^2} (\mathbf{r} \cdot \boldsymbol{\hat\sigma} - \mathbb{I}).
\end{equation}
The quantum Fisher information is defined as $F_Q = \mathrm{Tr}(\hat\varrho\hat L_\theta^2)$. Substituting the SLD and using the Pauli trace identities $\mathrm{Tr}(\hat\sigma^\alpha) = 0$ and $\mathrm{Tr}(\hat\sigma^\alpha \hat\sigma^\beta) = 2\delta^{\alpha\beta}$ reduces the expression to the standard exact formula~\cite{BraunsteinCaves1994}
\begin{equation}
F_j(t;\theta) = |\partial_\theta\mathbf{r}_j|^2 + \frac{(\mathbf{r}_j\cdot\partial_\theta\mathbf{r}_j)^2}{1-|\mathbf{r}_j|^2}.
\end{equation}
Because the second term is strictly non-negative, and both terms share the denominator bound $(1-|\mathbf{r}_j|^2)^{-1}$ upon taking a common denominator, the exact value is structurally bounded from below and above by the tangent norm
\begin{equation}
|\partial_\theta\mathbf{r}_j|^2 \le F_j(t;\theta) \le \frac{|\partial_\theta\mathbf{r}_j|^2}{1 - |\mathbf{r}_j|^2}.
\end{equation}
Multiplying the left inequality by $(1-|\mathbf{r}_j|^2)$ directly establishes the double-sided kinematic hierarchy utilized in the main text.

\section{Magnon-sector decomposition of the $U(1)$-breaking Hamiltonian}
\label{app:block_magnon}

This Appendix provides the golden-rule-like perturbative derivation supporting the state leakage analyzed in Sec.~\ref{sec:perturbation}. 

The analytical mapping of the local QFI onto the one-magnon pure-state wavepacket requires strict particle number conservation. Introducing the transverse-field perturbation
\begin{equation}
  \hat V = \sum_{i} \hat\sigma_i^x = \sum_i (\hat\sigma_i^+ + \hat\sigma_i^-)
\end{equation}
with coupling strength $h$ breaks this continuous $U(1)$ symmetry. This corresponds geometrically to decomposing the full Hamiltonian into a block-tridiagonal operator spanning orthogonal $n$-magnon submanifolds
\begin{equation}
\Hhat = \sum_{n=0}^{N}\hat P_n \Hhat_0 \hat P_n + h \sum_{n=0}^{N-1} \Big(\hat P_{n+1} \hat V\hat P_{n} +\hat P_{n}\hat V\hat P_{n+1} \Big),
\end{equation}
where $\hat P_n$ projects onto the $n$-magnon Hilbert subspace. 

The intra-manifold blocks along the main diagonal govern the non-interacting spatial hopping dynamics
\begin{equation}
  \hat P_n \Hhat_0\hat P_n =\hat P_n \left[ 2J \sum_i (\hat\sigma_i^+ \hat\sigma_{i+1}^- + \hat\sigma_{i}^- \hat\sigma_{i+1}^+) \right]\hat P_n.
\end{equation}
The off-diagonal inter-manifold ladder couplings are generated strictly by the external symmetry-breaking field
\begin{align}
  h\hat P_{n+1}\hat V\hat P_{n} &= h\hat P_{n+1} \Big( \sum_i \hat\sigma_i^- \Big)\hat P_{n}, \nonumber \\
  h\hat P_{n} \hat V\hat P_{n+1} &= h\hat P_{n} \Big( \sum_i \hat\sigma_i^+ \Big)\hat P_{n+1}.
\end{align}
These matrix couplings map directly to the phenomenological $\ket{\delta\phi_2(t)}$ two-magnon and $\ket{\delta\phi_0(t)}$ vacuum leakage channels isolated via the Dyson series expansion in Sec.~\ref{sec:perturbation}. We now formulate the corresponding FGR decay rate into the two-magnon continuum.

\subsection{Perturbative two-magnon leakage estimate}

In this subsection we give only a scaling estimate
for leakage from the one-magnon tangent channel
into the two-magnon sector.
The estimate is not intended to reproduce
finite-open-chain prefactors;
it only identifies the perturbative order in $h$
and the density-of-states scale set by $J$.

Under the Jordan--Wigner transformation,
the unperturbed XX Hamiltonian is diagonalized by the operators
\begin{equation}
  \hat{\tilde c}_k = \frac{1}{\sqrt{N}} \sum_\ell e^{-ik\ell}\hat c_\ell, \qquad \varepsilon_k = 4J\cos k.
\end{equation}
This transformation explicitly maps the $S=1/2$ spins to spinless fermions. Because fermions anticommute on different sites, the local spin flip operators are dressed by a non-local parity string to enforce canonical commutation relations. The local parity $\hat\sigma_j^z$ maps to the fermion occupation $\hat\sigma_j^z = 1 - 2\hat c_j^\dagger\hat c_j$, and the perturbation becomes the effective global operator
\begin{equation}
  h\hat V = h\sum_i (\hat c_i^\dagger +\hat c_i) \prod_{j<i} \hat\sigma_j^z.
\end{equation}
Although $\hat V$ is nonlocal, its action on a single-magnon basis state $\ket{\ell} =\hat c_\ell^\dagger\ket{\emptyset}$ yields an exact two-site operator action
\begin{equation}
  \hat\sigma_i^x \ket{\ell} = \prod_{j<i} (1 - 2\hat c_j^\dagger\hat c_j)(\hat c_i^\dagger +\hat c_i)\hat c_\ell^\dagger \ket{\emptyset}.
\end{equation}
If $i<\ell$, the string $\prod_{j<i} \hat\sigma_j^z$ encounters only empty sites. Because the reference vacuum is the fully polarized state $\ket{\psi_0} = \ket{\uparrow\dots\uparrow}$, any site without a fermion evaluates to local parity $+1$. The string therefore yields $+1$, giving
\begin{equation}
  \hat\sigma_i^x \ket{\ell} =\hat c_i^\dagger\hat c_\ell^\dagger \ket{\emptyset}.
\end{equation}
If $i>\ell$, the string operator evaluates to $-1$ because it overlaps with the single fermion at $\ell$. Applying the new fermion creation operator $\hat c_i^\dagger$ 
yields $- \hat c_i^\dagger\hat c_\ell^\dagger \ket{\emptyset}$. Because $i > \ell$, bringing this state back into the canonical spatially ordered form requires swapping the fermion operators. 
This swap introduces a second minus sign from fermionic anticommutation ($\{\hat c_i^\dagger,\hat c_\ell^\dagger\}=0$), exactly canceling the parity string
\begin{equation}
  \hat\sigma_i^x \ket{\ell} = -\hat c_i^\dagger\hat c_\ell^\dagger \ket{\emptyset} =\hat c_\ell^\dagger\hat c_i^\dagger \ket{\emptyset}.
\end{equation}
For $i \neq \ell$, we can summarize both cases
via the properly ordered two-magnon basis state
$\ket{a, b} = \hat c_a^\dagger\hat c_b^\dagger \ket{\emptyset}$ (with $a<b$) as
\begin{equation}
  \hat\sigma_i^x \ket{\ell} = \ket{\min(\ell, i), \max(\ell, i)}\qquad (i \neq \ell).
\end{equation}
For $i = \ell$, the operator annihilates the magnon
and returns the vacuum: $\hat\sigma_\ell^x\ket{\ell} = \ket{\emptyset}$.
Thus the projected action $\hat P_2\hat\sigma_i^x\ket{\ell}$
shows that all terms with $i \neq \ell$ promote
the one-magnon state into the two-magnon sector,
while the term $i = \ell$ contributes
to the vacuum channel $\hat P_0\hat V\hat P_1$.
The transition probability from a given one-magnon eigenstate
$\ket{\phi_1}$ into the two-magnon sector
follows from standard first-order time-dependent perturbation theory:
\begin{equation}
  P_{1\to 2}(t) = h^2 \sum_{\mu \in \mathcal{H}^{(2)}}|\langle \mu |\hat P_2\hat V\hat P_1 | \phi_1 \rangle|^2
  \frac{4\sin^2[(E_\mu - E_1)t/2]}{(E_\mu - E_1)^2}.
\end{equation}
At short times this grows as $h^2 t^2$.
In an intermediate regime where the two-magnon spectrum
can be treated as quasi-continuous,
the oscillatory kernel approaches $2\pi t\,\delta(E_\mu - E_1)$,
giving a golden-rule-like scaling
\begin{equation}
P_{1\to 2}(t) \sim \Gamma\, t,
\qquad
\Gamma \sim h^2\,\rho_2(E)\,\overline{|M|^2} \sim \frac{h^2}{J}\times(\text{dimensionless form factor}),
\end{equation}
where $\rho_2(E) \sim J^{-1}$ is the two-magnon density of states
set by the hopping energy scale.
This estimate is used only to motivate
the perturbative scaling of the leakage rate,
not to predict the finite open-chain numerical prefactor.
For the finite chains studied here ($N=10$--$20$),
the two-magnon sector is discrete
rather than a true continuum,
and finite-size recurrences limit the golden-rule regime.
The one-magnon survival probability depletes on a scale
$t \sim J/h^2$, consistent with the $\mathcal{O}(h^2 t/J)$
perturbative depletion derived in Sec.~\ref{sec:perturbation}.
The observed collapse of $\eta(t)/h^2$ in Fig.~\ref{fig:fgr_ratio}
verifies the perturbative order of the depletion
in the finite chains studied here.

\bibliographystyle{apsrev4-2}
\bibliography{references}

\begin{thebibliography}{28}%
\makeatletter
\providecommand \@ifxundefined [1]{%
 \@ifx{#1\undefined}
}%
\providecommand \@ifnum [1]{%
 \ifnum #1\expandafter \@firstoftwo
 \else \expandafter \@secondoftwo
 \fi
}%
\providecommand \@ifx [1]{%
 \ifx #1\expandafter \@firstoftwo
 \else \expandafter \@secondoftwo
 \fi
}%
\providecommand \natexlab [1]{#1}%
\providecommand \enquote  [1]{``#1''}%
\providecommand \bibnamefont  [1]{#1}%
\providecommand \bibfnamefont [1]{#1}%
\providecommand \citenamefont [1]{#1}%
\providecommand \href@noop [0]{\@secondoftwo}%
\providecommand \href [0]{\begingroup \@sanitize@url \@href}%
\providecommand \@href[1]{\@@startlink{#1}\@@href}%
\providecommand \@@href[1]{\endgroup#1\@@endlink}%
\providecommand \@sanitize@url [0]{\catcode `\\12\catcode `\$12\catcode
  `\&12\catcode `\#12\catcode `\^12\catcode `\_12\catcode `\%12\relax}%
\providecommand \@@startlink[1]{}%
\providecommand \@@endlink[0]{}%
\providecommand \url  [0]{\begingroup\@sanitize@url \@url }%
\providecommand \@url [1]{\endgroup\@href {#1}{\urlprefix }}%
\providecommand \urlprefix  [0]{URL }%
\providecommand \Eprint [0]{\href }%
\providecommand \doibase [0]{https://doi.org/}%
\providecommand \selectlanguage [0]{\@gobble}%
\providecommand \bibinfo  [0]{\@secondoftwo}%
\providecommand \bibfield  [0]{\@secondoftwo}%
\providecommand \translation [1]{[#1]}%
\providecommand \BibitemOpen [0]{}%
\providecommand \bibitemStop [0]{}%
\providecommand \bibitemNoStop [0]{.\EOS\space}%
\providecommand \EOS [0]{\spacefactor3000\relax}%
\providecommand \BibitemShut  [1]{\csname bibitem#1\endcsname}%
\let\auto@bib@innerbib\@empty
%</preamble>
\bibitem [{\citenamefont {Swingle}(2018)}]{Swingle2018}%
  \BibitemOpen
  \bibfield  {author} {\bibinfo {author} {\bibfnamefont {B.}~\bibnamefont
  {Swingle}},\ }\href {https://doi.org/10.1038/s41567-018-0295-5} {\bibfield
  {journal} {\bibinfo  {journal} {Nat. Phys.}\ }\textbf {\bibinfo {volume}
  {14}},\ \bibinfo {pages} {988} (\bibinfo {year} {2018})}\BibitemShut
  {NoStop}%
\bibitem [{\citenamefont {Xu}\ and\ \citenamefont
  {Swingle}(2024)}]{XuSwingle2024}%
  \BibitemOpen
  \bibfield  {author} {\bibinfo {author} {\bibfnamefont {S.}~\bibnamefont
  {Xu}}\ and\ \bibinfo {author} {\bibfnamefont {B.}~\bibnamefont {Swingle}},\
  }\href {https://doi.org/10.1103/PRXQuantum.5.010201} {\bibfield  {journal}
  {\bibinfo  {journal} {PRX Quantum}\ }\textbf {\bibinfo {volume} {5}},\
  \bibinfo {pages} {010201} (\bibinfo {year} {2024})}\BibitemShut {NoStop}%
\bibitem [{\citenamefont {Swingle}\ \emph {et~al.}(2016)\citenamefont
  {Swingle}, \citenamefont {Bentsen}, \citenamefont {Schleier-Smith},\ and\
  \citenamefont {Hayden}}]{Swingle2016}%
  \BibitemOpen
  \bibfield  {author} {\bibinfo {author} {\bibfnamefont {B.}~\bibnamefont
  {Swingle}}, \bibinfo {author} {\bibfnamefont {G.}~\bibnamefont {Bentsen}},
  \bibinfo {author} {\bibfnamefont {M.}~\bibnamefont {Schleier-Smith}},\ and\
  \bibinfo {author} {\bibfnamefont {P.}~\bibnamefont {Hayden}},\ }\href
  {https://doi.org/10.1103/PhysRevA.94.040302} {\bibfield  {journal} {\bibinfo
  {journal} {Phys. Rev. A}\ }\textbf {\bibinfo {volume} {94}},\ \bibinfo
  {pages} {040302(R)} (\bibinfo {year} {2016})}\BibitemShut {NoStop}%
\bibitem [{\citenamefont {G\"arttner}\ \emph {et~al.}(2017)\citenamefont
  {G\"arttner}, \citenamefont {Bohnet}, \citenamefont {Safavi-Naini},
  \citenamefont {Wall}, \citenamefont {Bollinger},\ and\ \citenamefont
  {Rey}}]{Garttner2017}%
  \BibitemOpen
  \bibfield  {author} {\bibinfo {author} {\bibfnamefont {M.}~\bibnamefont
  {G\"arttner}}, \bibinfo {author} {\bibfnamefont {J.~G.}\ \bibnamefont
  {Bohnet}}, \bibinfo {author} {\bibfnamefont {A.}~\bibnamefont
  {Safavi-Naini}}, \bibinfo {author} {\bibfnamefont {M.~L.}\ \bibnamefont
  {Wall}}, \bibinfo {author} {\bibfnamefont {J.~J.}\ \bibnamefont
  {Bollinger}},\ and\ \bibinfo {author} {\bibfnamefont {A.~M.}\ \bibnamefont
  {Rey}},\ }\href {https://doi.org/10.1038/nphys4119} {\bibfield  {journal}
  {\bibinfo  {journal} {Nat. Phys.}\ }\textbf {\bibinfo {volume} {13}},\
  \bibinfo {pages} {781} (\bibinfo {year} {2017})}\BibitemShut {NoStop}%
\bibitem [{\citenamefont {Xu}\ and\ \citenamefont
  {Swingle}(2020)}]{XuSwingle2020}%
  \BibitemOpen
  \bibfield  {author} {\bibinfo {author} {\bibfnamefont {S.}~\bibnamefont
  {Xu}}\ and\ \bibinfo {author} {\bibfnamefont {B.}~\bibnamefont {Swingle}},\
  }\href {https://doi.org/10.1038/s41567-019-0712-4} {\bibfield  {journal}
  {\bibinfo  {journal} {Nat. Phys.}\ }\textbf {\bibinfo {volume} {16}},\
  \bibinfo {pages} {199} (\bibinfo {year} {2020})}\BibitemShut {NoStop}%
\bibitem [{\citenamefont {Lieb}\ and\ \citenamefont
  {Robinson}(1972)}]{LiebRobinson1972}%
  \BibitemOpen
  \bibfield  {author} {\bibinfo {author} {\bibfnamefont {E.~H.}\ \bibnamefont
  {Lieb}}\ and\ \bibinfo {author} {\bibfnamefont {D.~W.}\ \bibnamefont
  {Robinson}},\ }\href {https://doi.org/10.1007/BF01645779} {\bibfield
  {journal} {\bibinfo  {journal} {Commun. Math. Phys.}\ }\textbf {\bibinfo
  {volume} {28}},\ \bibinfo {pages} {251} (\bibinfo {year} {1972})}\BibitemShut
  {NoStop}%
\bibitem [{\citenamefont {Braunstein}\ and\ \citenamefont
  {Caves}(1994)}]{BraunsteinCaves1994}%
  \BibitemOpen
  \bibfield  {author} {\bibinfo {author} {\bibfnamefont {S.~L.}\ \bibnamefont
  {Braunstein}}\ and\ \bibinfo {author} {\bibfnamefont {C.~M.}\ \bibnamefont
  {Caves}},\ }\href {https://doi.org/10.1103/PhysRevLett.72.3439} {\bibfield
  {journal} {\bibinfo  {journal} {Phys. Rev. Lett.}\ }\textbf {\bibinfo
  {volume} {72}},\ \bibinfo {pages} {3439} (\bibinfo {year}
  {1994})}\BibitemShut {NoStop}%
\bibitem [{\citenamefont {Hyllus}\ \emph {et~al.}(2012)\citenamefont {Hyllus},
  \citenamefont {Laskowski}, \citenamefont {Krischek}, \citenamefont
  {Schwemmer}, \citenamefont {Wieczorek}, \citenamefont {Weinfurter},
  \citenamefont {Pezz\`e},\ and\ \citenamefont {Smerzi}}]{Hyllus2012}%
  \BibitemOpen
  \bibfield  {author} {\bibinfo {author} {\bibfnamefont {P.}~\bibnamefont
  {Hyllus}}, \bibinfo {author} {\bibfnamefont {W.}~\bibnamefont {Laskowski}},
  \bibinfo {author} {\bibfnamefont {R.}~\bibnamefont {Krischek}}, \bibinfo
  {author} {\bibfnamefont {C.}~\bibnamefont {Schwemmer}}, \bibinfo {author}
  {\bibfnamefont {W.}~\bibnamefont {Wieczorek}}, \bibinfo {author}
  {\bibfnamefont {H.}~\bibnamefont {Weinfurter}}, \bibinfo {author}
  {\bibfnamefont {L.}~\bibnamefont {Pezz\`e}},\ and\ \bibinfo {author}
  {\bibfnamefont {A.}~\bibnamefont {Smerzi}},\ }\href
  {https://doi.org/10.1103/PhysRevA.85.022321} {\bibfield  {journal} {\bibinfo
  {journal} {Phys. Rev. A}\ }\textbf {\bibinfo {volume} {85}},\ \bibinfo
  {pages} {022321} (\bibinfo {year} {2012})}\BibitemShut {NoStop}%
\bibitem [{\citenamefont {T\'oth}(2012)}]{Toth2012}%
  \BibitemOpen
  \bibfield  {author} {\bibinfo {author} {\bibfnamefont {G.}~\bibnamefont
  {T\'oth}},\ }\href {https://doi.org/10.1103/PhysRevA.85.022322} {\bibfield
  {journal} {\bibinfo  {journal} {Phys. Rev. A}\ }\textbf {\bibinfo {volume}
  {85}},\ \bibinfo {pages} {022322} (\bibinfo {year} {2012})}\BibitemShut
  {NoStop}%
\bibitem [{\citenamefont {Pezz\`e}\ \emph {et~al.}(2018)\citenamefont
  {Pezz\`e}, \citenamefont {Smerzi}, \citenamefont {Oberthaler}, \citenamefont
  {Schmied},\ and\ \citenamefont {Treutlein}}]{Pezze2018}%
  \BibitemOpen
  \bibfield  {author} {\bibinfo {author} {\bibfnamefont {L.}~\bibnamefont
  {Pezz\`e}}, \bibinfo {author} {\bibfnamefont {A.}~\bibnamefont {Smerzi}},
  \bibinfo {author} {\bibfnamefont {M.~K.}\ \bibnamefont {Oberthaler}},
  \bibinfo {author} {\bibfnamefont {R.}~\bibnamefont {Schmied}},\ and\ \bibinfo
  {author} {\bibfnamefont {P.}~\bibnamefont {Treutlein}},\ }\href
  {https://doi.org/10.1103/RevModPhys.90.035005} {\bibfield  {journal}
  {\bibinfo  {journal} {Rev. Mod. Phys.}\ }\textbf {\bibinfo {volume} {90}},\
  \bibinfo {pages} {035005} (\bibinfo {year} {2018})}\BibitemShut {NoStop}%
\bibitem [{\citenamefont {Yoshida}\ and\ \citenamefont
  {Yao}(2019)}]{YoshidaYao2019}%
  \BibitemOpen
  \bibfield  {author} {\bibinfo {author} {\bibfnamefont {B.}~\bibnamefont
  {Yoshida}}\ and\ \bibinfo {author} {\bibfnamefont {N.~Y.}\ \bibnamefont
  {Yao}},\ }\href {https://doi.org/10.1103/PhysRevX.9.011006} {\bibfield
  {journal} {\bibinfo  {journal} {Phys. Rev. X}\ }\textbf {\bibinfo {volume}
  {9}},\ \bibinfo {pages} {011006} (\bibinfo {year} {2019})}\BibitemShut
  {NoStop}%
\bibitem [{\citenamefont {Lewis-Swan}\ \emph {et~al.}(2019)\citenamefont
  {Lewis-Swan}, \citenamefont {Safavi-Naini}, \citenamefont {Bollinger},\ and\
  \citenamefont {Rey}}]{LewisSwan2019}%
  \BibitemOpen
  \bibfield  {author} {\bibinfo {author} {\bibfnamefont {R.~J.}\ \bibnamefont
  {Lewis-Swan}}, \bibinfo {author} {\bibfnamefont {A.}~\bibnamefont
  {Safavi-Naini}}, \bibinfo {author} {\bibfnamefont {J.~J.}\ \bibnamefont
  {Bollinger}},\ and\ \bibinfo {author} {\bibfnamefont {A.~M.}\ \bibnamefont
  {Rey}},\ }\href {https://doi.org/10.1038/s41467-019-09436-y} {\bibfield
  {journal} {\bibinfo  {journal} {Nat. Commun.}\ }\textbf {\bibinfo {volume}
  {10}},\ \bibinfo {pages} {1581} (\bibinfo {year} {2019})}\BibitemShut
  {NoStop}%
\bibitem [{\citenamefont {G\"arttner}\ \emph {et~al.}(2018)\citenamefont
  {G\"arttner}, \citenamefont {Hauke},\ and\ \citenamefont
  {Rey}}]{Garttner2018}%
  \BibitemOpen
  \bibfield  {author} {\bibinfo {author} {\bibfnamefont {M.}~\bibnamefont
  {G\"arttner}}, \bibinfo {author} {\bibfnamefont {P.}~\bibnamefont {Hauke}},\
  and\ \bibinfo {author} {\bibfnamefont {A.~M.}\ \bibnamefont {Rey}},\ }\href
  {https://doi.org/10.1103/PhysRevLett.120.040402} {\bibfield  {journal}
  {\bibinfo  {journal} {Phys. Rev. Lett.}\ }\textbf {\bibinfo {volume} {120}},\
  \bibinfo {pages} {040402} (\bibinfo {year} {2018})},\ \Eprint
  {https://arxiv.org/abs/1706.01616} {arXiv:1706.01616 [quant-ph]} \BibitemShut
  {NoStop}%
\bibitem [{\citenamefont {Paris}(2009)}]{Paris2009}%
  \BibitemOpen
  \bibfield  {author} {\bibinfo {author} {\bibfnamefont {M.~G.~A.}\
  \bibnamefont {Paris}},\ }\href {https://doi.org/10.1142/S0219749909004839}
  {\bibfield  {journal} {\bibinfo  {journal} {Int. J. Quantum Inf.}\ }\textbf
  {\bibinfo {volume} {7}},\ \bibinfo {pages} {125} (\bibinfo {year}
  {2009})}\BibitemShut {NoStop}%
\bibitem [{\citenamefont {Liu}\ \emph {et~al.}(2020)\citenamefont {Liu},
  \citenamefont {Yuan}, \citenamefont {Lu},\ and\ \citenamefont
  {Wang}}]{Liu2020JPA}%
  \BibitemOpen
  \bibfield  {author} {\bibinfo {author} {\bibfnamefont {J.}~\bibnamefont
  {Liu}}, \bibinfo {author} {\bibfnamefont {H.}~\bibnamefont {Yuan}}, \bibinfo
  {author} {\bibfnamefont {X.-M.}\ \bibnamefont {Lu}},\ and\ \bibinfo {author}
  {\bibfnamefont {X.}~\bibnamefont {Wang}},\ }\href
  {https://doi.org/10.1088/1751-8121/ab5d4d} {\bibfield  {journal} {\bibinfo
  {journal} {J. Phys. A: Math. Theor.}\ }\textbf {\bibinfo {volume} {53}},\
  \bibinfo {pages} {023001} (\bibinfo {year} {2020})}\BibitemShut {NoStop}%
\bibitem [{\citenamefont {Roberts}\ and\ \citenamefont
  {Swingle}(2016)}]{RobertsSwingle2016}%
  \BibitemOpen
  \bibfield  {author} {\bibinfo {author} {\bibfnamefont {D.~A.}\ \bibnamefont
  {Roberts}}\ and\ \bibinfo {author} {\bibfnamefont {B.}~\bibnamefont
  {Swingle}},\ }\href {https://doi.org/10.1103/PhysRevLett.117.091602}
  {\bibfield  {journal} {\bibinfo  {journal} {Phys. Rev. Lett.}\ }\textbf
  {\bibinfo {volume} {117}},\ \bibinfo {pages} {091602} (\bibinfo {year}
  {2016})}\BibitemShut {NoStop}%
\bibitem [{\citenamefont {Xu}\ and\ \citenamefont
  {Swingle}(2019)}]{XuSwingle2019}%
  \BibitemOpen
  \bibfield  {author} {\bibinfo {author} {\bibfnamefont {S.}~\bibnamefont
  {Xu}}\ and\ \bibinfo {author} {\bibfnamefont {B.}~\bibnamefont {Swingle}},\
  }\href {https://doi.org/10.1103/PhysRevX.9.031048} {\bibfield  {journal}
  {\bibinfo  {journal} {Phys. Rev. X}\ }\textbf {\bibinfo {volume} {9}},\
  \bibinfo {pages} {031048} (\bibinfo {year} {2019})}\BibitemShut {NoStop}%
\bibitem [{\citenamefont {Lieb}\ \emph {et~al.}(1961)\citenamefont {Lieb},
  \citenamefont {Schultz},\ and\ \citenamefont
  {Mattis}}]{LiebSchultzMattis1961}%
  \BibitemOpen
  \bibfield  {author} {\bibinfo {author} {\bibfnamefont {E.}~\bibnamefont
  {Lieb}}, \bibinfo {author} {\bibfnamefont {T.}~\bibnamefont {Schultz}},\ and\
  \bibinfo {author} {\bibfnamefont {D.}~\bibnamefont {Mattis}},\ }\href
  {https://doi.org/10.1016/0003-4916(61)90115-4} {\bibfield  {journal}
  {\bibinfo  {journal} {Ann. Phys.}\ }\textbf {\bibinfo {volume} {16}},\
  \bibinfo {pages} {407} (\bibinfo {year} {1961})}\BibitemShut {NoStop}%
\bibitem [{\citenamefont {Lin}\ and\ \citenamefont
  {Motrunich}(2018)}]{Lin2018}%
  \BibitemOpen
  \bibfield  {author} {\bibinfo {author} {\bibfnamefont {C.-J.}\ \bibnamefont
  {Lin}}\ and\ \bibinfo {author} {\bibfnamefont {O.~I.}\ \bibnamefont
  {Motrunich}},\ }\href {https://doi.org/10.1103/PhysRevB.97.144304} {\bibfield
   {journal} {\bibinfo  {journal} {Phys. Rev. B}\ }\textbf {\bibinfo {volume}
  {97}},\ \bibinfo {pages} {144304} (\bibinfo {year} {2018})}\BibitemShut
  {NoStop}%
\bibitem [{\citenamefont {Lopez-Piqueres}\ \emph {et~al.}(2021)\citenamefont
  {Lopez-Piqueres}, \citenamefont {Ware}, \citenamefont {Gopalakrishnan},\ and\
  \citenamefont {Vasseur}}]{LopezPiqueres2021}%
  \BibitemOpen
  \bibfield  {author} {\bibinfo {author} {\bibfnamefont {J.}~\bibnamefont
  {Lopez-Piqueres}}, \bibinfo {author} {\bibfnamefont {B.}~\bibnamefont
  {Ware}}, \bibinfo {author} {\bibfnamefont {S.}~\bibnamefont
  {Gopalakrishnan}},\ and\ \bibinfo {author} {\bibfnamefont {R.}~\bibnamefont
  {Vasseur}},\ }\href {https://doi.org/10.1103/PhysRevB.104.104307} {\bibfield
  {journal} {\bibinfo  {journal} {Phys. Rev. B}\ }\textbf {\bibinfo {volume}
  {104}},\ \bibinfo {pages} {104307} (\bibinfo {year} {2021})}\BibitemShut
  {NoStop}%
\bibitem [{\citenamefont {Colmenarez}\ and\ \citenamefont
  {Luitz}(2020)}]{Colmenarez2020}%
  \BibitemOpen
  \bibfield  {author} {\bibinfo {author} {\bibfnamefont {L.}~\bibnamefont
  {Colmenarez}}\ and\ \bibinfo {author} {\bibfnamefont {D.~J.}\ \bibnamefont
  {Luitz}},\ }\href {https://doi.org/10.1103/PhysRevResearch.2.043047}
  {\bibfield  {journal} {\bibinfo  {journal} {Phys. Rev. Research}\ }\textbf
  {\bibinfo {volume} {2}},\ \bibinfo {pages} {043047} (\bibinfo {year}
  {2020})}\BibitemShut {NoStop}%
\bibitem [{\citenamefont {Xu}\ \emph {et~al.}(2020)\citenamefont {Xu},
  \citenamefont {Scaffidi},\ and\ \citenamefont {Cao}}]{XuScaffidi2020}%
  \BibitemOpen
  \bibfield  {author} {\bibinfo {author} {\bibfnamefont {T.}~\bibnamefont
  {Xu}}, \bibinfo {author} {\bibfnamefont {T.}~\bibnamefont {Scaffidi}},\ and\
  \bibinfo {author} {\bibfnamefont {X.}~\bibnamefont {Cao}},\ }\href
  {https://doi.org/10.1103/PhysRevLett.124.140602} {\bibfield  {journal}
  {\bibinfo  {journal} {Phys. Rev. Lett.}\ }\textbf {\bibinfo {volume} {124}},\
  \bibinfo {pages} {140602} (\bibinfo {year} {2020})},\ \Eprint
  {https://arxiv.org/abs/1912.11063} {arXiv:1912.11063 [cond-mat.stat-mech]}
  \BibitemShut {NoStop}%
\bibitem [{\citenamefont {Wysocki}\ and\ \citenamefont
  {Chwede\ifmmode~\acute{n}\else \'{n}\fi{}czuk}(2025)}]{Wysocki2025}%
  \BibitemOpen
  \bibfield  {author} {\bibinfo {author} {\bibfnamefont {P.}~\bibnamefont
  {Wysocki}}\ and\ \bibinfo {author} {\bibfnamefont {J.}~\bibnamefont
  {Chwede\ifmmode~\acute{n}\else \'{n}\fi{}czuk}},\ }\href
  {https://doi.org/10.1103/PhysRevLett.134.020201} {\bibfield  {journal}
  {\bibinfo  {journal} {Phys. Rev. Lett.}\ }\textbf {\bibinfo {volume} {134}},\
  \bibinfo {pages} {020201} (\bibinfo {year} {2025})}\BibitemShut {NoStop}%
\bibitem [{\citenamefont {{\v S}afr{\'a}nek}(2017)}]{Safranek2017}%
  \BibitemOpen
  \bibfield  {author} {\bibinfo {author} {\bibfnamefont {D.}~\bibnamefont {{\v
  S}afr{\'a}nek}},\ }\href {https://doi.org/10.1103/PhysRevA.95.052320}
  {\bibfield  {journal} {\bibinfo  {journal} {Phys. Rev. A}\ }\textbf {\bibinfo
  {volume} {95}},\ \bibinfo {pages} {052320} (\bibinfo {year}
  {2017})}\BibitemShut {NoStop}%
\bibitem [{\citenamefont {B{\"{a}}rtschi}\ and\ \citenamefont
  {Eidenbenz}(2019)}]{Bartschi2019}%
  \BibitemOpen
  \bibfield  {author} {\bibinfo {author} {\bibfnamefont {A.}~\bibnamefont
  {B{\"{a}}rtschi}}\ and\ \bibinfo {author} {\bibfnamefont {S.}~\bibnamefont
  {Eidenbenz}},\ }in\ \href {https://doi.org/10.1007/978-3-030-25027-0_9}
  {\emph {\bibinfo {booktitle} {Fundamentals of Computation Theory}}}\
  (\bibinfo  {publisher} {Springer International Publishing},\ \bibinfo {year}
  {2019})\ pp.\ \bibinfo {pages} {126--139}\BibitemShut {NoStop}%
\bibitem [{\citenamefont {Yang}(2025)}]{Yang2025}%
  \BibitemOpen
  \bibfield  {author} {\bibinfo {author} {\bibfnamefont {Y.}~\bibnamefont
  {Yang}},\ }\href {https://doi.org/10.1103/PhysRevLett.134.010603} {\bibfield
  {journal} {\bibinfo  {journal} {Phys. Rev. Lett.}\ }\textbf {\bibinfo
  {volume} {134}},\ \bibinfo {pages} {010603} (\bibinfo {year}
  {2025})}\BibitemShut {NoStop}%
\bibitem [{\citenamefont {Ferro}\ and\ \citenamefont
  {Fagotti}(2025)}]{FerroFagotti2025}%
  \BibitemOpen
  \bibfield  {author} {\bibinfo {author} {\bibfnamefont {F.}~\bibnamefont
  {Ferro}}\ and\ \bibinfo {author} {\bibfnamefont {M.}~\bibnamefont
  {Fagotti}},\ }\href@noop {} {\bibfield  {journal} {\bibinfo  {journal}
  {arXiv:2503.21905}\ } (\bibinfo {year} {2025})},\ \bibinfo {note}
  {preprint},\ \Eprint {https://arxiv.org/abs/2503.21905} {arXiv:2503.21905
  [quant-ph]} \BibitemShut {NoStop}%
\bibitem [{\citenamefont {Hayden}\ and\ \citenamefont
  {Preskill}(2007)}]{HaydenPreskill2007}%
  \BibitemOpen
  \bibfield  {author} {\bibinfo {author} {\bibfnamefont {P.}~\bibnamefont
  {Hayden}}\ and\ \bibinfo {author} {\bibfnamefont {J.}~\bibnamefont
  {Preskill}},\ }\href {https://doi.org/10.1088/1126-6708/2007/09/120}
  {\bibfield  {journal} {\bibinfo  {journal} {J. High Energy Phys.}\ }\textbf
  {\bibinfo {volume} {2007}}\bibinfo  {number} { (09)},\ \bibinfo {pages}
  {120}}\BibitemShut {NoStop}%
\end{thebibliography}%
\end{document}